\documentclass{dhbenelux}

\usepackage{booktabs} %
\usepackage{authblk} %
\usepackage{graphicx} %

\usepackage{url}
\usepackage{hyperref}
\usepackage{lineno}
\usepackage{microtype}
\usepackage{subcaption}
\usepackage{algpseudocode,algorithm,algorithmicx}
\usepackage{empheq}
\usepackage{xr}
\usepackage{amssymb}

\newcommand{\bn}{{\mathbf{n}}}

\newcommand{\bA}{{\mathbf{A}}}

\newcommand{\bw}{{\mathbf{w}}}

\newcommand{\bx}{{\mathbf{x}}}
\newcommand{\bh}{{\mathbf{h}}}
\newcommand{\bP}{{\mathbf{P}}}
\newcommand{\bQ}{{\mathbf{Q}}}

\newcommand{\bk}{{\mathbf{k}}}
\newcommand{\bI}{{\mathbf{I}}}

\newcommand{\bD}{{\mathbf{D}}}

\newcommand{\bW}{{\mathbf{W}}}

\DeclareMathOperator{\sinc}{sinc}

\author[1]{Kathleen MacWilliam}
\author[1]{Thomas Dietzen}
\author[1,2]{Randall Ali}
\author[1]{Toon van Waterschoot}

\affil[1]{STADIUS, Department of Electrical Engineering (ESAT), KU Leuven, Leuven, Belgium}
\affil[2]{Institute of Sound Recording, Department of Music and Media, University of Surrey, Guildford, United Kingdom}

\title{State-Space Estimation of Spatially Dynamic Room Impulse Responses using a Room Acoustic Model-based Prior}

\begin{document}

\maketitle

\begin{abstract}
    \noindent {The estimation of room impulse responses (RIRs) between static loudspeaker and microphone locations can be done using a number of well-established measurement and inference procedures. While these procedures assume a time-invariant acoustic system, time variations need to be considered for the case of spatially dynamic scenarios where loudspeakers and microphones are subject to movement. If the RIR is modeled using image sources, then movement implies that the distance to each image source varies over time, making the estimation of the spatially dynamic RIR particularly challenging. In this paper, we propose a procedure to estimate the early part of the spatially dynamic RIR between a stationary source and a microphone moving on a linear trajectory at constant velocity. The procedure is built upon a state-space model, where the state to be estimated represents the early RIR, the observation corresponds to a microphone recording in a spatially dynamic scenario, and time-varying distances to the image sources are incorporated into the state transition matrix obtained from static RIRs at the start and end point of the trajectory. The performance of the proposed approach is evaluated against state-of-the-art RIR interpolation and state-space estimation methods using simulations, demonstrating the potential of the proposed state-space model.}
\end{abstract}


\section{Introduction}
A room impulse response (RIR) is the time-domain representation of the linear time-invariant (LTI) system that uniquely characterizes the cumulative impact of a room on sound waves between a specific static source and microphone position, effectively representing the room's acoustic environment. This concept is fundamental to various acoustic signal processing applications, including source localization \citep{evers2020locata}, dereverberation \citep{naylor2010speech}, echo cancellation \citep{elko2003room}, and spatial audio reproduction \citep{schissler2017efficient}, among others. Considerable research efforts have been dedicated to developing robust measurement \citep{stan2002comparison,Szke2018BuildingAE}  and estimation \citep{Ratnarajah2022TowardsIR, Crocco2015RoomIR, 1597551} techniques for capturing RIRs, particularly in scenarios where the source and microphone remain static within the acoustic environment \citep{stan2002comparison,Szke2018BuildingAE}. However, real-world situations often involve spatially dynamic scenarios, where sources and microphones are subject to movement. This paper addresses the challenge of accurately estimating the early part of RIRs along a trajectory in a time-varying acoustic scenario with a stationary source and a moving microphone. This can equivalently be thought of as a time-varying system identification problem, where the system to be identified may be referred to as a \textit{time-varying} RIR. 

In this context, it is important to elaborate on the apparent contradiction between the previously defined RIR as an LTI system representation and the concept of a time-variant. 
In this paper, the acoustic environment itself is assumed to be time-invariant, in which case an RIR between two static positions indeed corresponds to an LTI system. 
RIRs, however, are inherently location-variant, i.e. they depend on the locations of the source and the microphone. 
The \textit{time-varying} RIR, as considered in this paper, can be defined as the time-varying system representation \citep{cherniakov2003introduction} relating the source signal to the microphone signal if the microphone moves, i.e. if the microphone has a time-varying location. 
In discrete time, this time-variant RIR can also be thought of as a collection of RIRs in the LTI sense over different time-instances: the location of the microphone changes at each time step, leading to a new discrete position in space, associated with a time-invariant RIR. 
Time-variant RIR estimation, as defined above, has relevance across numerous acoustic signal processing applications, especially amid the growing interest in virtual acoustic environments. For instance, \cite{ajdler2007dynamic} demonstrated the use of time-varying acoustic system models for enabling rapid measurements of head-related impulse responses. Moreover, sub-optimal estimations of time-variant RIRs can impair the effectiveness of echo-cancellation systems in telepresence and communication technologies \citep{Nophut2024}.

In order to contextualize our contribution in the estimation of the early part of a time-variant RIR, it is necessary to provide a brief outline of the various related literature. 
In this context, the concept of spatial RIR interpolation is highly relevant given that we can consider a time-variant RIR as a collection of RIRs over a discrete set of locations.
RIR interpolation facilitates sound field rendering for dynamic source-microphone positions by filling spatial gaps in RIR data, as measurements or simulations are usually limited to sparse grids. Numerous approaches have been proposed for RIR interpolation including compressed sensing methods \citep{mignot2013room,8521390}, spherical harmonics \citep{borra2019soundfield}, physics-based models \citep{antonello2017room, Hahmann2022ACP}, directional RIRs \citep{zhao2022interpolating}, and neural networks \citep{perini2021deep, Karakonstantis2024RoomIR}. \cite{haneda1999common} introduced a frequency-domain approach for interpolating room transfer functions (RTFs), particularly effective at lower frequencies, later extended by \cite{das2021room}. Additionally, several techniques for the interpolation of head-related transfer functions (HRTFs), which are critical for accurate binaural rendering and share some methodological approaches with RTF interpolation, have been explored, as discussed in \cite{carty2010movements}.
For the purpose of this paper, we focus on the interpolation of an RIR at a point between two microphone locations given their estimated RIRs for a common stationary source. \cite{kearney2009dynamic} introduced Dynamic Time Warping (DTW)-based interpolation, dividing RIRs into early reflections and diffuse decay. This method temporally aligns and linearly interpolates early reflections while modeling the tail using \cite{masterson2009acoustic}'s approach, laying the foundation for subsequent developments. \cite{garcia2018binaural} expanded on Kearney's work, enhancing algorithm robustness and computational efficiency while retaining the core interpolation technique. Building on this, \cite{bruschi2020innovative} refined peak finding and matching aspects of the interpolation technique. \cite{geldert2023interpolation} proposed a novel approach using partial optimal transport for interpolation, enabling non-bijective mapping of sound events between the early part of RIRs. 
It is important to highlight that these interpolation approaches operate outside of a conventional system identification framework. Instead, they use a limited set of measured RIRs and predominantly rely on a room acoustic sound propagation model such as the image source method (ISM) \citep{allen1979image}.
As opposed to interpolation strategies, fully data-driven approaches for estimating time-variant RIRs have also been investigated. In this case, the estimation relies directly on the source and microphone signals and can be framed as a system identification or adaptive filtering problem. %
This approach has been largely motivated by the need to obtain rapid measurements of head related impulse responses \citep{Hahn2015,Enzner2008} and for echo-cancellation \citep{Antweiler1995, enzner2010bayesian}, where a typical scenario involves an excitation signal being continuously captured by a moving microphone. Using carefully designed excitation signals \citep{Hahn2015, Kuhl2018}, an estimation of the time-variant RIR can be done using a Normalized Least Mean Squares (NLMS) algorithm \citep{Enzner2008, Antweiler2012} or more generally using a Kalman filter as in \cite{enzner2010bayesian}. 
In the context of this work, a Kalman filter is of particular interest as it is derived from a state-space model of the dynamic system, where the state to be estimated is the time-variant RIR \citep{enzner2010bayesian}. 
In such a state-space model, the evolution of the time-variant RIR is explicitly modeled by means of a first-order difference equation, which allows more modeling flexibility than other adaptive algorithms suchs as NLMS.
One popular choice for the first-order difference model is to relate the states at two time instants by a transition factor and an additional process noise term. 
The transition coefficient and process noise covariance are typically set according to the expected variability of the state, influencing the convergence behavior of the Kalman filter.
For instance, in \cite{Nophut2024}, the transition coefficient was modeled as a function of the microphone velocity.

In this paper, we consider the problem of estimating the early part of the time-variant RIR between a stationary source and a microphone moving on a linear trajectory at constant velocity. We propose integrating RIR interpolation, derived from a room acoustic model, into a state-space-based framework for RIR estimation, thereby merging data-driven approaches with physical modeling. More specifically, rather than relying solely on a state transition factor within the state equation, we propose incorporating the ISM into a state transition matrix between the early segments of consecutive RIRs. We derive an analytical model for this transition matrix and subsequently propose estimating it from static RIRs at the trajectory's start and end points using a DTW-based algorithm.
The proposed approach's performance is evaluated through simulations by comparing it to two alternatives: one that relies solely on the state equation for RIR estimation, resembling a pure interpolation method with the ISM-based transition matrix, and another that uses a conventional state-space estimation with a simple state transition factor. Our findings suggest that the proposed state-space model outperforms both alternatives in terms of normalized misalignment between the simulated `ground-truth' RIRs and the estimated RIRs.

The subsequent sections of this paper are organized as follows: Section \ref{problem_statement_SOTA} introduces the signal model and provides an overview of the most pertinent state-of-the-art methods. Section \ref{proposedapproach} elaborates on the proposed RIR state-space model and outlines the update equations of the Kalman filter used to recursively estimate the defined state. Section \ref{transitionmatrices} offers detailed derivations of the proposed room acoustic model-based state transition matrix. Finally, Section \ref{simulations} presents experimental validation through simulations, followed by a discussion of the results. 

\section{Signal Model, Problem Statement, and Related State-of-the-Art}
\label{problem_statement_SOTA}
\subsection{Signal model and Problem Statement}
\begin{figure}
\begin{center}
\includegraphics[width=12cm]{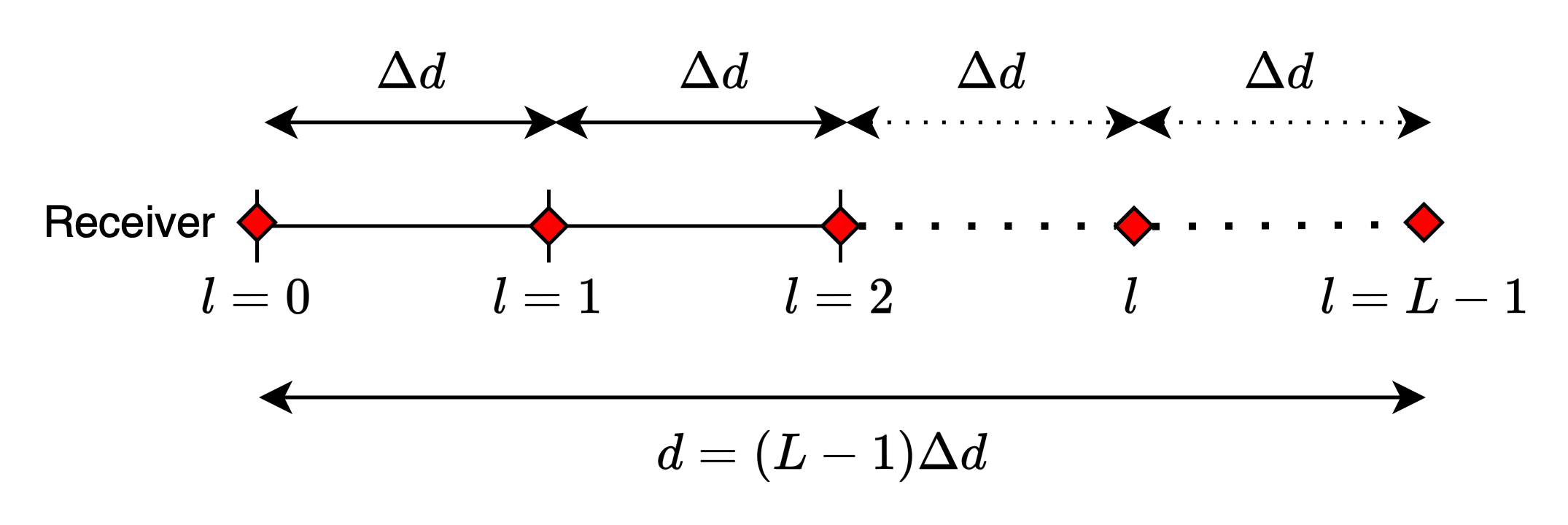}%
\end{center}
\caption{Linear microphone trajectory of length $d$ divided into $L$ equidistant locations spaced by $\Delta d$ and indexed as $l = 0,\,\dots,\,L-1$ such that $d = (L-1)\Delta d$.}
\label{fig:traj1}
\end{figure}

\begin{figure}
\begin{center}
\includegraphics[width=13cm]{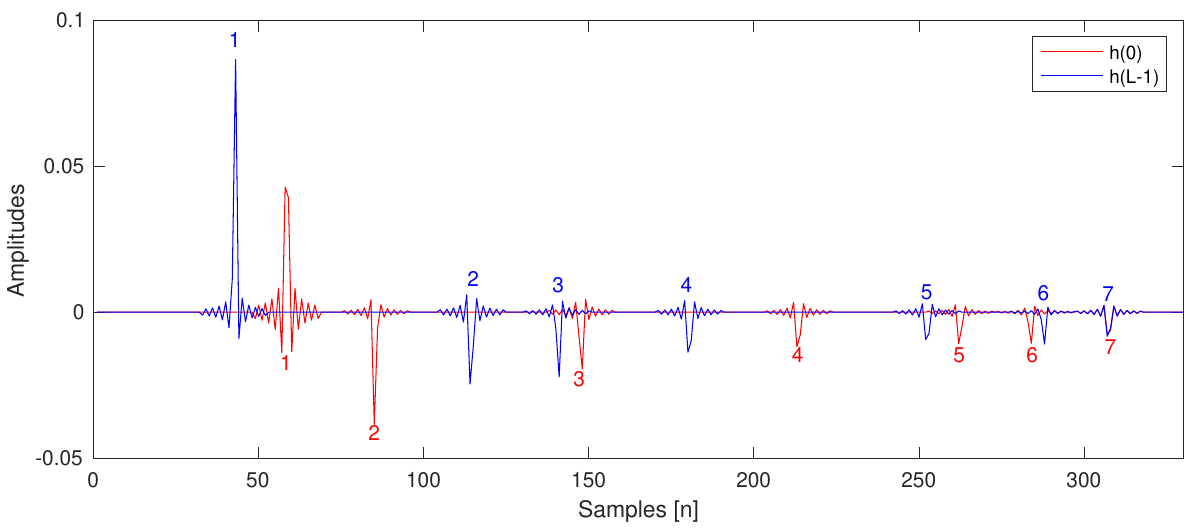}%
\end{center}
\caption{An example of simulated RIRs, $\bh(0)$ and $\bh(L-1)$, including only first-order reflections. Peaks corresponding to the same source or reflection are labeled by the same number in each RIR.}\label{fig:impulse}
\end{figure}

We firstly introduce the signal model to  formally define the concept of an RIR as employed in this paper.  In the remainder of this paper, when the term ``RIR'' is mentioned, it will specifically pertain to the early part of the RIRs, as we do not address the estimation of the late reverberant tail. We also assume that the source location remains static.
Let $k$ denote the time index, and let the location of the microphone relative to the linear trajectory be denoted by the one-dimensional location index $l$.
If the microphone location is time-variant, an RIR characterizes the LTI system relating a source signal $x(k)$ to an observed signal $y(l, k)$. 
Let the RIR for microphone location $l$ be denoted by $h(l,n)$, where $n$ indexes the time-shift of the RIR samples. 
The signals $x(k)$ and $y(l,k)$ are then related by  
\begin{equation}
 y(l,k)= \sum_{n=0}^{N-1} h(l,n)x(k-n) + v(k),
   \label{definitionRIR}
\end{equation}
where $v(k)$ is a noise term that we generally assume to be present in the observed signal and may include late reverberation. Now, defining the vectors $\bx(k)$ and $\bh(l)$ as
\begin{align}
\bx(k) &= \begin{pmatrix}
x(k) & x(k-1) & \dots & x(k - N + 1)
\end{pmatrix}^T,\\
\bh(l) &= \begin{pmatrix}
h(l,0) & h(l,1) & \dots & h(l,N-1)
\end{pmatrix}^T,\label{h_vec}
\end{align}
the convolution in \eqref{definitionRIR} can alternatively be written as
\begin{align}
 y(l,k) &=\bx^T(k)\bh(l) + v(k).
   \label{definitionRIR_vec}
\end{align}
Within the scope of this paper, we assume a linear microphone trajectory of length $d$, divided into $L$ equidistant locations spaced by $\Delta d$ and indexed as $l = 0,\,\dots,\,L-1$, as depicted in Figure \ref{fig:traj1}, such that $L = d/\Delta d + 1$.
We consider the problem of estimating $\bh(l)$ using the source signal $x(k)$ and the observed signal $y(l, k)$ if the microphone moves along the trajectory at a constant velocity $\mu_{\operatorname{rx}}$, i.e. if $l$ varies over time.
For this, we make the assumptions that $\bh(0)$ and $\bh(L-1)$ are known, and that the time intervals which the individual reflections occupy within the time-variant RIR over the range of the trajectory do not overlap. An example of $\bh(0)$ and $\bh(L-1)$ including only first-order reflections is shown in Figure \ref{fig:impulse}, where peaks corresponding to the same source or reflection are labeled by the same number in each RIR.

\subsection{Related State-of-the-Art}
\label{section_SOTA}
Before introducing the proposed approach, it is instructive to briefly introduce the most relevant concepts used in the state of the art. 
On the one hand, we consider RIR interpolation approaches that estimate $\bh(l)$ from $\bh(0)$ and $\bh(L-1)$ only based on a room acoustic sound propagation model, without making use of an observed signal $y(l, k)$. 
On the other hand, we consider data-driven approaches that estimate $\bh(l)$ in an adaptive manner from a data set containing $x(k)$ and $y(l, k)$ over multiple time indexes $k$ for a moving microphone, but without exploiting a room acoustic sound propagation model. 

As previously mentioned, some RIR interpolation approaches \citep{kearney2009dynamic,geldert2023interpolation} are motivated by the ISM for room acoustic sound propagation, which expresses an RIR as a sum of contributions from the original source and so-called image sources representing reflections from the boundaries of the room. 
Based on this model, RIR interpolation approaches infer the location-variant time of arrival (TOA), as well as the amplitude of the direct component and individual reflections (or equivalently, source and image source components) at a particular location $l$ using the RIRs $\bh(0)$ and $\bh(L-1)$.
The general principle can be described as follows. 
Assuming that reflections corresponding to the same image source
in $\bh(0)$ and $\bh(L-1)$ can be identified, let $\bn_r(0)$ and $\bn_r(L-1)$ be vectors collecting the respective time-shift indices of the direct component and these reflections in $\bh(0)$ and $\bh(L-1)$. 
The corresponding time-shift indices of $\bh(l)$ can be interpolated as
\begin{equation}
\bn_r(l) = \bn_r(0) + \operatorname{round}\Bigl(\beta(l) \bigl(\bn_r(L-1) - \bn_r(0)\bigr)\Bigr),
\label{sample_based_interpolation}
\end{equation}
where $\bn_r(L-1) - \bn_r(0)$ can be thought of as sample-based time differences of arrival (TDOAs), 
$\beta(l) \in [0,1]$  with $\beta(0) = 0$ and $\beta(L-1) = 1$, and $\operatorname{round}(\cdot)$ defines a rounding operation.
The function $\beta(l)$ can be defined such that the interpolation is correct for the direct component \citep{kearney2009dynamic}.
Applying a similar interpolation rule for the amplitudes of the direct component and reflections, an estimate of $\bh(l)$ can then be obtained.
The collections of corresponding indices $\bn_r(0)$ and $\bn_r(L-1)$ can for instance be found by means of dynamic time warping with $\bh(0)$ and $\bh(L-1)$ as inputs \citep{kearney2009dynamic,bruschi2020innovative}.
Here, an accurate identification of all corresponding reflections requires that their order of arrival is the same in $\bh(0)$ and $\bh(L-1)$.

Data-driven approaches \citep{enzner2010bayesian} to the estimation of  location- or time-variant RIRs do not rely on explicit modeling of room acoustic sound propagation, but instead perform adaptive system identification given a data set containing $x(k)$ and $y(l, k)$ over multiple time indexes $k$.
For the moment, let us assume that the moving microphone crosses one location $l$ at each time instant $k$, such that the location-variant RIR $\bh(l)$ can equivalently be thought of as a time-variant RIR $\bh(k)$ with $k=l$. 
The convolution in \eqref{definitionRIR}
then becomes
\begin{align}
 y(k)&= \sum_{n=0}^{N-1} h(k,n)x(k-n) + v(k).
 \label{definitionTimeVaryingRIR}
\end{align}
A popular approach to adaptively estimate $\bh(k)$ is the Kalman filter, whose update equations are obtained from a state-space model where $\bh(k)$ is the state.
A commonly used state-space model can be defined as
\begin{align}
 \bh(k) &= \alpha\bh(k-1) + \bw(k),\label{state_equation_sota}\\
 y(k) &=\bx^T(k)\bh(k) + v(k),
   \label{observation_equation_sota}
\end{align}
where the state equation in \eqref{state_equation_sota} models the evolution of the state as a first-order difference equation, and the observation equation in \eqref{observation_equation_sota} relates the state to the observation $y(k)$ and is identical to \eqref{definitionTimeVaryingRIR}.
The model commonly chosen in \eqref{state_equation_sota} defines a factor $\alpha$, referred to as the forgetting factor or transition factor, and a noise term $\bw(k)$, referred to as process noise, for which the covariance matrix is required in the Kalman filter. In practice, the choice of $\alpha$ and the covariance matrix of $\bw(k)$ is typically not directly motivated by a physical model of the RIR evolution. 
Instead, they are often considered tuning parameters that can be used to control the convergence behavior of the Kalman filter, although they can be tuned depending on physical parameters such as the velocity of the microphone \citep{Nophut2024}.

\section{Proposed RIR State-space Model}\label{proposedapproach}
In the proposed approach, we aim to include prior knowledge on room acoustic sound propagation into a state-space model for time-variant RIR estimation.
To this end, rather than resorting to a 
scalar factor as in \eqref{state_equation_sota}, we assume that the relation between two RIRs $\bh(l)$ and $\bh(l-1)$ on the trajectory can approximately be modeled by a location-invariant transition matrix $\bA$ as
\begin{equation}
 \bh(l) \approx \bA \bh(l-1)
 \label{tranConst}
\end{equation}
for $l=1,\,\dots,\,L-1$. The transition matrix $\bA$ will be used to model changes in the TOA of the direct component and the early reflections between the neighbouring locations $l$ and $l-1$ and will be applied in the state equation of the proposed state-space model. As described in Section \ref{transitionmatrices}, the analytical definition of $\bA$ will be based on the ISM, which also serves as the foundation of the RIR interpolation approaches discussed in Section \ref{section_SOTA}.
Furthermore, we will show that the transition matrix $\bA$ can be modeled as location-invariant if the time intervals which the individual reflections occupy in the RIR over the range of the trajectory do not overlap, which is implicitly also assumed in  \citep{kearney2009dynamic,geldert2023interpolation}.
In practice, an estimate of $\bA$ can be obtained from the presumed knowledge of the RIRs $\bh(0)$ and $\bh(L-1)$ at the start and the end of the trajectory, as will be discussed in Section \ref{dtw}.

While it is possible to interpolate between $\bh(0)$ and $\bh(L-1)$ based on $\bA$ only as $\bh(l) \approx \bA^l \bh(0)$, a more accurate estimate can be obtained by using the signals recorded along the trajectory with a  microphone moving at a constant velocity. 
Without loss of generality, 
let $k=0$ denote the start of the recording at location $l=0$.
For a given velocity $\mu_{\operatorname{rx}}$ of the microphone and temporal sampling period $T_s$, the smallest possible spatial period $\Delta d$ that could be used is given by $\mu_{\operatorname{rx}}T_s$. In this case, one RIR estimate can be obtained per time sample.
In the proposed approach, we also consider spatial periods of 
\begin{equation}
\Delta d = \Omega \mu_{\operatorname{rx}}T_s,
\label{spatialPeriod}
\end{equation}
where $\Omega$ is a positive integer that can be understood as a spatial downsampling factor. 
If the microphone is at location with index $l$, we then observe sample 
\begin{align}
k &= l \Omega.
\label{k2n}
\end{align}
With \eqref{tranConst} and defining $y_\Omega(l) = y(l\Omega)$,  $\bx_\Omega(l) = \bx(l\Omega)$, and $v_\Omega(l) = v(l\Omega)$ for ease of presentation, we can define the state-space model of the proposed approach as
\begin{empheq}[box=\fbox]{align}
    \bh(l) &= \bA\bh(l-1) + \bw(l),\label{state_equation_prop}\\
    y_\Omega(l) &= \bx_\Omega^T(l)\bh(l) + v_\Omega(l), \label{observation_equation_prop}
\end{empheq}
where $\bw(l)$ is the process noise accounting for modeling errors in the definition of $\bA$.
From the state-space model, an estimate $\hat{\bh}(l)$ of ${\bh}(l)$ can be obtained using the Kalman filter with one recursion per location $l$. The Kalman filter update equations are outlined in the following Sec. \ref{sec_update_equations}.
\subsection{Update Equations of the Kalman Filter}\label{sec_update_equations}
The Kalman filter \citep{simon2006optimal} can be used to recursively estimate the state defined by a state-space model.
For the proposed state-space model in \eqref{state_equation_prop}-\eqref{observation_equation_prop}, let $\bQ = \operatorname{E}[\bw(l)\bw^T(l)]$ and $R = \operatorname{E}[v_{\Omega}^2(l)]$ be the presumed location-invariant covariance matrix of $\bw(l)$ and variance of $v_{\Omega}(l)$, respectively, where $\operatorname{E}[\cdot]$ denotes the expectation operation.
The update equations of the Kalman filter for the model in \eqref{state_equation_prop}-\eqref{observation_equation_prop} are then given by
\begin{align}
    \hat{\bh}(l) & = \bA \hat{\bh}^{+}(l-1), \label{eq:state-prediction} \\
    \bP(l) & = \bA \bP^{+}(l-1) \bA^T + \bQ, \label{eq:covariance-prediction} \\[5pt]
    \bk(l) & = \dfrac{\bP(l) \bx_\Omega(l)}{\bx_\Omega^T(l) \bP(l) \bx_\Omega(l) + R}, \label{eq:kalman-gain} \\
    \hat{\bh}^{+}(l) & = \hat{\bh}(l) + \bk(l) \left(y_{\Omega}(l) - \bx_{\Omega}^T(l) \hat{\bh}(l)\right), \label{eq:state-update} \\
    \bP^{+}(l) & = \left(\bI - \bk(l) \bx_\Omega^T(l)\right) \bP(l), \label{eq:covariance-update}
\end{align}
where \eqref{eq:state-prediction}--\eqref{eq:covariance-prediction} are commonly referred to as the prediction step producing prior estimates, and \eqref{eq:kalman-gain}--\eqref{eq:covariance-update} are referred to as the update step producing posterior estimates.
In these equations, $\hat{\bh}(l)$ and $\hat{\bh}^+(l)$ are state estimates, $\bP(l)$ and $\bP^+(l)$ are estimates of the state-estimation error covariance matrix, $\bk(l)$ is the Kalman gain, $\bI$ the identity matrix, and the superscript $^+$ distinguishes posterior from prior estimates.
In the prediction step \eqref{eq:state-prediction}--\eqref{eq:covariance-prediction}, the previously acquired posterior estimates $\hat{\bh}^+(l-1)$ and $\bP^+(l-1)$ are propagated based on the state equation \eqref{state_equation_prop} only, i.e. without taking into account the observation at recursion $l$, yielding the prior estimates $\hat{\bh}(l)$ and $\bP(l)$.
The update step \eqref{eq:kalman-gain}--\eqref{eq:covariance-update} is based on the observation equation \eqref{observation_equation_prop}. Here, the Kalman gain $\bk(l)$ and the error term $y_{\Omega}(l) - \bx_{\Omega}^T(l) \hat{\bh}(l)$ are computed based on $y_{\Omega}(l)$ and $\bx_{\Omega}(l)$ and used to update the prior estimates $\hat{\bh}(l)$ and $\bP(l)$, yielding the posterior estimates $\hat{\bh}^+(l)$ and $\bP^+(l)$.

To implement the Kalman filter, $\bQ$ and $R$ are required as inputs.
While $R$ can potentially be obtained from background noise recordings, $\bQ$ can be considered a tuning parameter as the variance of $\bw(l) = {\bh}(l) - \bA{\bh}(l-1)$ will not be known exactly in practice. 
The Kalman filter also  needs to be initialized by defining $\hat{\bh}^{+}(0)$ and $\bP^{+}(0)$.
As we assume that ${\bh}(0)$ is known, we set $\hat{\bh}^{+}(0) = {\bh}(0)$. This choice of $\hat{\bh}^{+}(0)$ corresponds to the assumption that the initial state estimation error is zero, such that $\bP^{+}(0)$ can be chosen to have a small norm. 
Finally, a practical implementation requires an estimate of $\bA$, which will be discussed in more detail in Section \ref{locationInvAnalytical} and \ref{dtw}.

At this point, it is instructive to interpret \eqref{eq:state-prediction}--
\eqref{eq:covariance-update} in relation to the state of the art as discussed in Section \ref{problem_statement_SOTA}.
In the interpolation approaches in \citep{kearney2009dynamic,geldert2023interpolation} on the one hand, recorded signals are not available, which corresponds to the assumption that $y_{\Omega}(l) = 0$ and $\bx_{\Omega}(l) = 0$.
In this case, we find that $\hat{\bh}^{+}(l) = \hat{\bh}(l)$ and $\bP^{+}(l) = \bP(l)$ in \eqref{eq:state-update}-\eqref{eq:covariance-update}, i.e. $\hat{\bh}^{+}(l)$ does not depend on $\bP(l)$ anymore and the Kalman filter update equations effectively reduce to $\hat{\bh}(l) = \bA \hat{\bh}(l-1) = \bA^l\bh(0)$. 
As $\bA$ is defined by interpolating TOAs of reflections between ${\bh}(0)$ and ${\bh}(L-1)$ based on a room acoustic model, this case is conceptually similar to the aforementioned interpolation approaches and can be referred to as linear interpolation.
The state-space space approaches in \cite{enzner2010bayesian} on the other hand are obtained if $\Omega = 1$ and the state transition matrix $\bA$ is replaced by a scalar $\alpha$.

\section{Proposed Room Acoustic Model-based Transition Matrix}\label{transitionmatrices}
This section provides a detailed explanation of the methodology followed to obtain a suitable room acoustic model-based transition matrix for use in Equation \eqref{state_equation_prop} and is organized as follows. In Section \ref{analy} we derive an analytical expression for a location-variant transition matrix model $\bA(l)$ based on the ISM. A suitable modification of the analytical transition matrix for use in the state-space model is then proposed in Section \ref{locationInvAnalytical} in order to obtain a location-invariant transition matrix model $\bA$ valid within the limits of the trajectory. The presented models are parameterized by the exact reflection TOAs, which are not inherently available in practice. Therefore, Section \ref{dtw} introduces an approach based on DTW to obtain an estimate of $\bA$ from $\bh(0)$ and $\bh(L-1)$ by TDOA interpolation.
\subsection{Analytical Location-Variant Transition Matrix Model}\label{analy}
\begin{figure}
\begin{center}
\includegraphics[width=12cm]{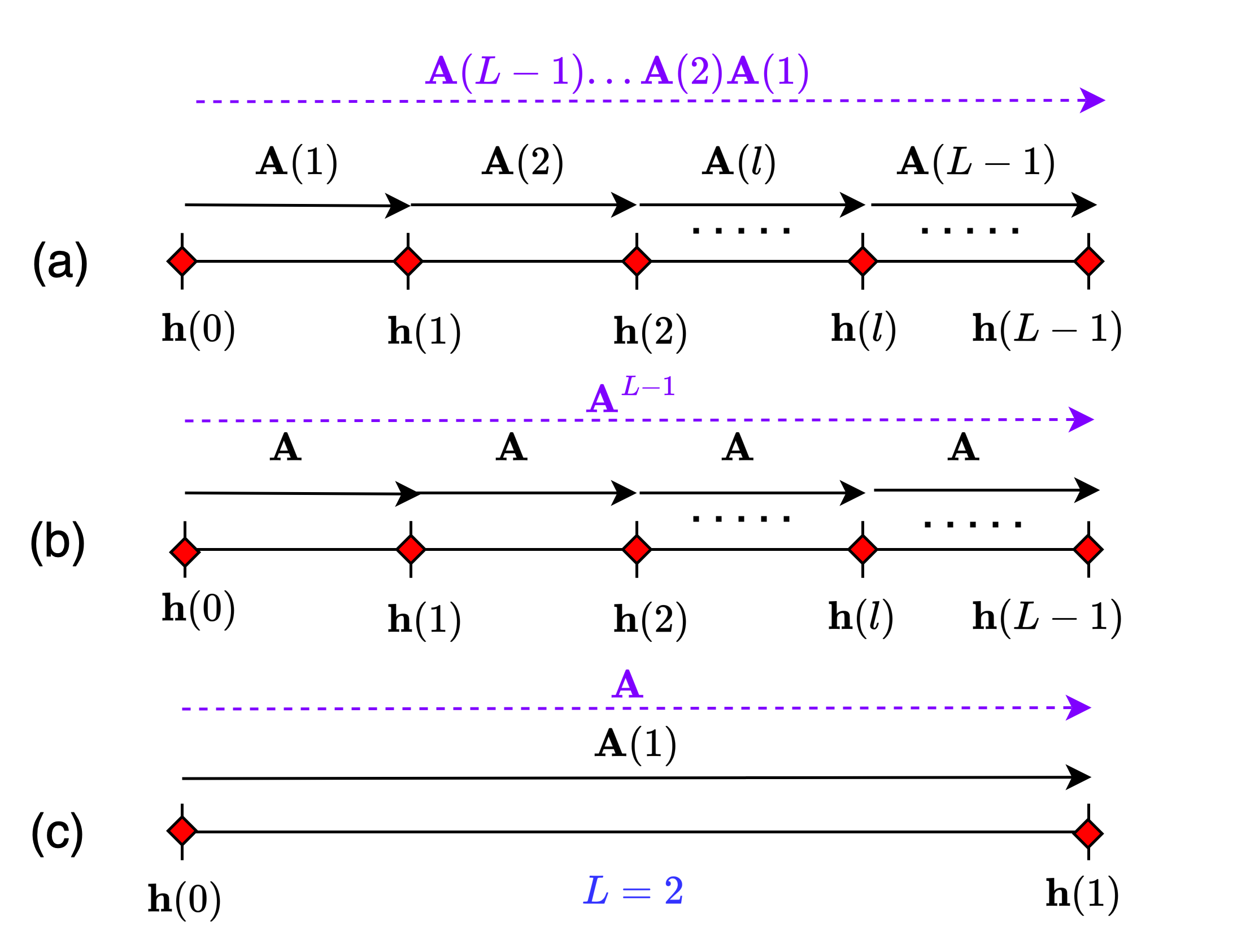}%
\end{center}
\caption{\textbf{(a)} Location-variant transition matrix model $\bA(l)$ providing mapping between two RIRs $\bh(l)$ and $\bh(l-1)$ along a given trajectory. \textbf{(b)} Location-invariant transition matrix model $\bA$ that approximates the mapping between two RIRs $\bh(l)$ and $\bh(l-1)$ along a given trajectory where the assumptions detailed in Section \ref{locationInvAnalytical} hold. \textbf{(c)} Equivalent transition matrix models, $\bA(1)$ and $\bA$, for the special case of $L=2$, i.e. a transition matrix between the start and end points of the trajectory.}
\label{fig:traj2}
\end{figure}

\begin{figure}[ht]
\begin{center}
\includegraphics[width=13cm]{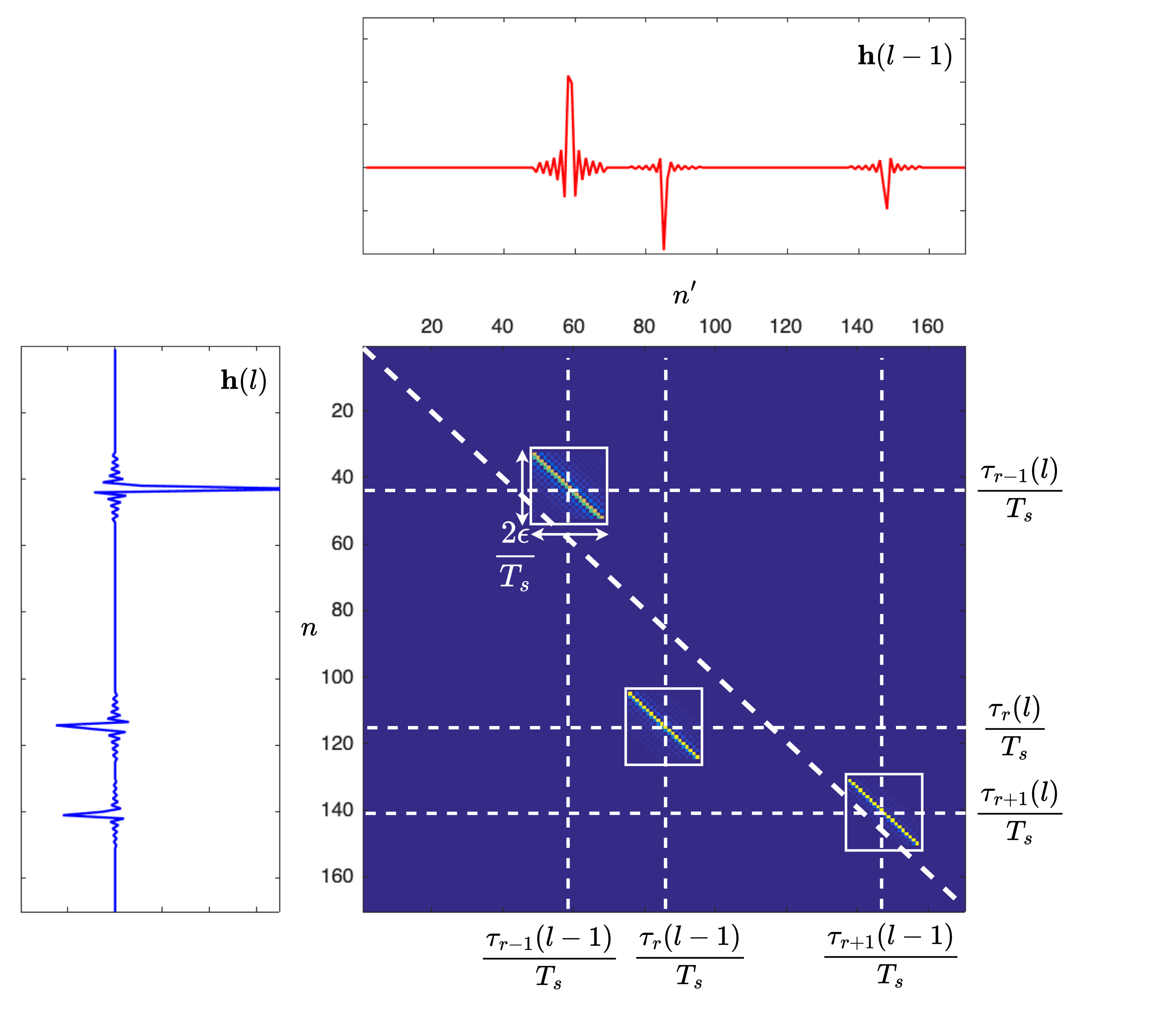}
\end{center}
\caption{An example of an analytical location-variant transition matrix $\bA(l)$ constructed from the early segments of two impulse responses, $\bh(l)$ and $\bh(l-1)$. The grid lines indicate the positions of the respective reflection TOAs}
\label{fig:A(l)}
\end{figure}

The objective of this section is to derive a transition matrix model $\bA(l) \in \mathbb{R}^{N \times N}$ that approximates the mapping between two  RIRs $\bh(l)$ and $\bh(l-1)$, i.e.
\begin{equation}
 \bh(l) \approx \bA(l) \bh(l-1).
 \label{tran}
\end{equation}
An illustration of this relation along a given trajectory is shown in Figure \ref{fig:traj2} (a).
To achieve this objective, we make use of the ISM. In this model, the RIR is expressed as the sum of contributions from the original sound source and additional image sources, which represent reflections within the room. 
Let the RIR at location $l$ with continuous time-shift $t$ be denoted by $h(l, t)$. With  $\delta(t)$ denoting the Dirac delta function, the ISM models $h(l, t)$ as
\begin{equation}
   h(l, t)=  \sum_{r \in \mathcal{R}} a_{r}(l) \delta  ( t-\tau_{r}(l)),
   \label{ISM}
\end{equation}
where $r \in \mathcal{R}=\{0,1,..R-1\}$ is the (image) source or reflection index, and $\tau_{r}(l)$ and $a_{r}(l)$ respectively denote the TOA and amplitude of the (image) source $r$ observed at location $l$.
Ultimately, we aim to express a discretized version of $h(l, t)$ in terms of a discretized version of $h(l-1, t)$ as in \eqref{tran}. To this end, we start by introducing parameters of $h(l-1,t)$ into $h(l,t)$ as follows,
\begin{equation}
 h(l,t)=  \sum_{r\in\mathcal{R}} a_{r}(l-1) \frac{a_{r}(l)}{a_{r}(l-1)} \delta \bigl( t-\tau_{r}(l) +\tau_{r}(l-1)-\tau_{r}(l-1)\bigr).\label{ism_mod}
\end{equation}
The inclusion of $a_{r}(l-1)/a_{r}(l-1)$ and $\tau_{r}(l-1)-\tau_{r}(l-1)$ preserves the definition of $h(l,t)$ in \eqref{ISM}, while allowing us to find a representation that more closely resembles $h(l-1,t)$. By incorporating $\tau_{r}(l-1)-\tau_{r}(l-1)$, we can isolate a Dirac delta function that only includes a time shift of $\tau_{r}(l-1)$ using the following identity derived from the sifting property \citep{hoskins2009delta} 
\begin{equation}
\delta \bigl( t-\tau_{r}(l) +\tau_{r}(l-1)- \tau_{r}(l-1)\bigr)=\int^{+\infty}_{-\infty} \delta (t-\tau_{r}(l) +\tau_{r}(l-1)-t')\delta (t'-\tau_{r}(l-1)\bigr) dt' 
\end{equation}
Therefore, \eqref{ism_mod} can alternatively be expressed as
\begin{equation}
h(l,t) =  \sum_{r\in\mathcal{R}} \int_{t'}  \frac{a_{r}(l)}{a_{r}(l-1)} \delta \bigl( t- \tau_{r}(l)+ \tau_{r}(l-1)-t'\bigr) a_{r}(l-1) \delta \bigl(t'-\tau_{r}(l-1)\bigr)\,dt'.
 \label{conti}
\end{equation}
where $t'$ is the integration range.
To obtain a discrete time-shift representation in terms of the time-shift index $n$, we impose a bandlimit with critical sampling period \(T_s\). Bandlimiting a Dirac delta function results in a sinc-function \citep{hoskins2009delta}, which will be sampled at discrete time indices. Defining 
\begin{align}
\Delta_{r}(l) =\tau_{r}(l)-\tau_{r}(l-1)
\label{TDOA}
\end{align}
as the TDOA of reflection $r$ between locations $l$ and $l-1$ we obtain
\begin{equation}
 h(l,n) =  \sum_{r\in \mathcal{R}} \sum_{n'} \frac{a_{r}(l)}{a_{r}(l-1)} \sinc \left( n- \frac{\Delta_{r}(l)}{T_s}-n' \right)  a_{r}(l-1) \sinc \left(n'-\frac{\tau_{r}(l-1)}{T_s}\right)
 \label{h2}.
\end{equation}
At this point, it is advantageous to introduce a time-shift dependent approximation of \eqref{h2}. At any $n$, the  RIR sample $ h(l,n)$ is dominated by a subset of reflections only, which we denote as $\tilde{\mathcal{R}}(l,n)$.
Due to the decaying nature of the sinc-function, the product $\sinc \left( n- {\Delta_{r}(l)}/{T_s}-n' \right)\sinc \left(n'-{\tau_{r}(l-1)}/T_s\right)$ in \eqref{h2} will have a prominent peak only if the two sinc-functions are closely aligned, i.e. if $nT_s$ is near ${\Delta_{r}(l)} + {\tau_{r}(l-1)} = {\tau_{r}(l)}$.
In the following, the product is considered negligible for $|nT_s-{\tau_{r}(l)}| > \epsilon$, where $\epsilon$ is a misalignment threshold and $2\epsilon$ can be understood as the effective temporal width of reflection $r$.
For ease of presentation, we define the interval of width $2\epsilon$ around ${\tau_{r}(l)}$ as
\begin{align}
\mathcal{T}_r(l) = [\tau_{r}(l)-\epsilon,\, \tau_{r}(l)+\epsilon],
\label{interval_Trl}
\end{align}
which contains the TOAs of the non-negligible components of reflection $r$ in the RIR at location $l$.
With \eqref{interval_Trl}, the misalignment condition $|nT_s-{\tau_{r}(l)}| > \epsilon$ can alternatively be expressed as $n \not\in \mathcal{T}_r(l)/T_s$.
Based on this, we define $\tilde{\mathcal{R}}(l,n)$ as \footnote{For notational convenience when working with an interval \([a, b]\), we introduce \(\frac{[a, b]}{c}\) to represent \([\frac{a}{c}, \frac{b}{c}]\) (see \eqref{subset} and \eqref{subset2}), and \([a, b] - c\) to represent \([a - c, b - c]\) (see \eqref{interval_Trmin1}).}
 \begin{equation}
\boxed{\tilde{\mathcal{R}}(l,n) =\left\{r \in \mathcal{R} \,\left|\, n \in \frac{\mathcal{T}_r(l)}{T_s} \right.\right\}.}
 \label{subset}
 \end{equation}
With \eqref{subset}, we can therefore approximate \eqref{h2} as
\begin{equation}
 h(l,n) \approx \sum_{n'} \sum_{r\in \tilde{\mathcal{R}}(l,n)}  \frac{a_{r}(l)}{a_{r}(l-1)} \sinc \left( n- \frac{\Delta_{r}(l)}{T_s}-n' \right)  a_{r}(l-1) \sinc \left(n'-\frac{\tau_{r}(l-1)}{T_s}\right),
 \label{h2approx}
\end{equation}
where we have swapped the order of summation over $n$ and $r$. Analogous to \eqref{ISM}, we introduce the coefficients $h(l-1, n')$ defined by the ISM as
\begin{equation}
h(l-1, n') = \sum_{r' \in \mathcal{R} } a_{r'}(l-1) \sinc\left (n'-\frac{\tau_{r'}(l-1)}{T_s}\right)
\label{ISMdiscrete}
\end{equation}
and rewrite \eqref{h2approx} as
\begin{equation}
 h(l,n) \approx \sum_{n'} \sum_{r\in \tilde{\mathcal{R}}(l,n)}  \frac{a_{r}(l)}{a_{r}(l-1)} \sinc \left( n- \frac{\Delta_{r}(l)}{T_s}-n' \right)  h(l-1, n') - e(l,n)
 \label{h2h1}
\end{equation}
with the error term $e(l,n)$ defined as
\begin{equation}
e(l,n) = \sum_{n'}
\sum_{\substack{r \in \tilde{\mathcal{R}}(l,n)\\ r' \in \mathcal{R} \\ r' \neq r}} 
\frac{a_{r}(l)}{a_{r}(l-1)}\sinc\left( n- \frac{\Delta_{r}(l)}{T_s}-n'\right)a_{r'}(l-1)\sinc\left (n'-\frac{\tau_{r'}(l-1)}{T_s}\right).
 \label{err}
\end{equation}
Given the relation between  $h(l,n)$ and $ h(l-1,n')$ in \eqref{h2h1}, we can define the elements of $\bA(l)$ in \eqref{tran} assuming that $e(l,n)$ is negligible.
Before turning to the elements of $\bA(l)$, let us first verify under which conditions this assumption holds. Similarly to before, we can argue that the product $\sinc( n- {\Delta_{r}(l)}/{T_s}-n')\sinc (n'-{\tau_{r'}(l-1)}/{T_s})$
will be negligible if the two sinc-functions are misaligned, i.e. if $|nT_s-{\Delta_{r}(l)} - {\tau_{r'}(l-1)}| > \epsilon$. 
As shown in Appendix A, this condition is equivalent to
\begin{equation}
\mathcal{T}_{r'}(l-1) \cap  \mathcal{T}_{r}(l-1) = \varnothing,  \quad r' \neq r,  
\label{assumption2}
\end{equation}
which essentially says that the reflections in $h(l-1,n)$ are required to be well-separated in time. 
Neglecting $e(l,n)$ and referring back to the vector representation of RIRs in \eqref{h_vec}, it can be seen that \eqref{h2h1} can be expressed in the form $\bh(l) \approx \bA(l) \bh(l-1)$ as anticipated in \eqref{tran}, where the element of $\bA(l)$ at index $(n+1, n'+1)$ is given by
\begin{equation}
\boxed{A_{n+1,n'+1}(l) = \sum_{r\in \tilde{\mathcal{R}}(l,n)}  \frac{a_{r}(l)}{a_{r}(l-1)} \sinc \left( n- \frac{\Delta_{r}(l)}{T_s}-n' \right)}
\label{elements_transition_matrix}
\end{equation}
if $\tilde{\mathcal{R}}(l,n) \neq \varnothing$ and otherwise $A_{n+1,n'+1}(l) = 0$, i.e. at indices where $h(l,n)$ does not contain reflections.
In the actual implementation of such a matrix, one may  choose to truncate the sinc-functions in equation \eqref{elements_transition_matrix} for $|n'T_s - \tau_r(l-1)| > \epsilon$.
As illustrated in Figure \ref{fig:A(l)}, this results in a matrix $\bA(l)$ composed of Toeplitz-structured submatrices of size $\lfloor \frac{2\epsilon}{ T_s} \rfloor \times \lfloor \frac{2\epsilon}{T_s} \rfloor$ containing sampled sinc-functions, with one such submatrix per reflection $r$. 
These sampled sinc-functions are vertically and horizontally centered around $\tau_r(l)/T_s$ and $\tau_{r}(l-1)/T_s$, which are non-integer in general.\footnote{In the special case that $\tau_r(l)/T_s$ and $\tau_{r}(l-1)/T_s$ are integer, the sinc-functions are sampled symmetrically around their peak, which implies that all but the peak sample lie at the zero-crossings of the sinc-functions, resulting in diagonal submatrices.} 
Practically speaking, we note that submatrices centered above the main diagonal of $\bA(l)$ shift the corresponding reflection in $\bh(l-1)$ towards smaller TOAs, while entries below shift it towards larger TOAs.

\subsection{Analytical Location-Invariant Transition Matrix Model}
\label{locationInvAnalytical}
For the application at hand, we anticipate employing a location-invariant transition matrix model as described in \eqref{tranConst} and illustrated in Figure \ref{fig:traj2} (b). We show that by using a location-invariant matrix, in which TOA intervals are defined to span the entire trajectory, we can reduce the necessity for accurate TOA estimates. %
This efficiency is achieved because these TOA intervals, along with TDOA estimates, can be obtained directly from $\bh(0)$ to $\bh(L-1)$ using the DTW-based method outlined in Section \ref{dtw}. In this Section, we therefore derive a location-invariant transition matrix $\bA$ from the definition of $\bA(l)$ in \eqref{elements_transition_matrix}.
As will be shown, location-invariance can be achieved within the limits of the trajectory under some assumptions.

\begin{figure}
\begin{center}
\includegraphics[width=14cm]{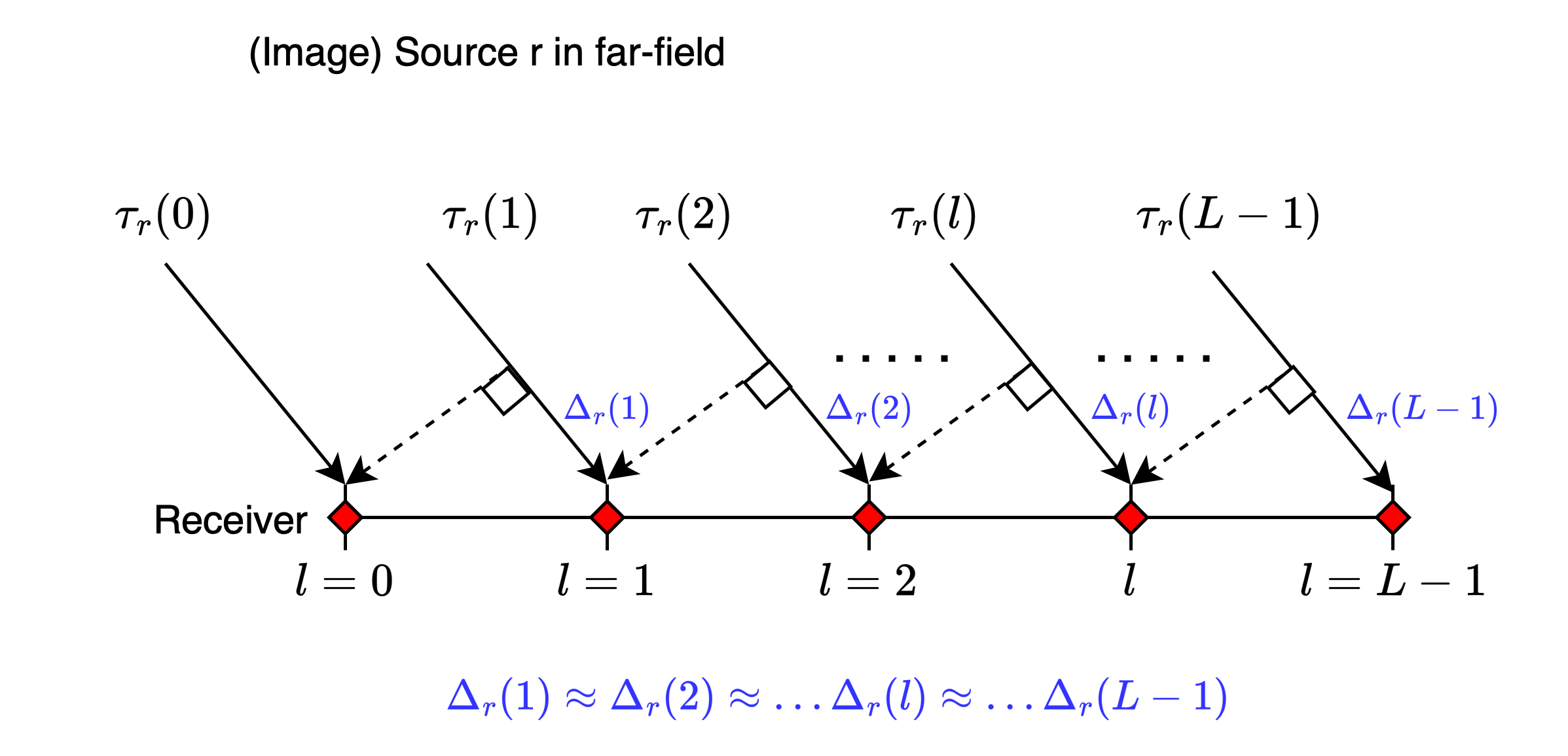}%
\end{center}
\caption{Illustration of the far-field assumption used to mitigate the dependence of ${\Delta}_r(l)$ on $l$.  }\label{fig:farfield} 
\end{figure}

In \eqref{elements_transition_matrix}, we observe two dependencies on $l$, namely in the TDOA ${\Delta}_r(l)$ and in the set $\tilde{\mathcal{R}}(l,n)$.
The dependency of ${\Delta}_r(l)$ on $l$ disappears under the following conditions. As illustrated in Figure \ref{fig:farfield}, under the assumptions of a far-field scenario where the (image) sources are far away in relation to the length of the trajectory,  we have that ${a_{r}(l)}/{a_{r}(l-1)} = 1$ and ${\Delta}_r(l) = {\Delta}_r$, meaning corresponding reflection TOAs are shifted by the same amount at each step along the trajectory. 
Consequently, the reflection TOA ${\tau}_r(l)$ can be approximated by the linear relation
\begin{equation}
    {\tau}_r(l) = {\tau}_r(0) +  l {\Delta}_r.
    \label{linearTOAs}
\end{equation}
The dependency of $\tilde{\mathcal{R}}(l,n)$ on $l$ as defined in \eqref{subset} can be resolved as follows. 
The model in \eqref{tranConst} is required to hold for $l=1,\,\dots,\,L-1$. 
Over this range of $l$, the components of reflection $r$ occupy the interval
\begin{align}
 \mathcal{T}_r &= [{\tau}_{r|\operatorname{min}},\, {\tau}_{r|\operatorname{max}}]  \label{interval_Tr1}\\
 \text{with} \quad {\tau}_{r|\operatorname{min}} &= \operatorname{min}\bigl({\tau}_r(1),\, {\tau}_r(L-1) \bigr) - \epsilon,  \label{interval_Tr2}\\
 {\tau}_{r|\operatorname{max}} &= \operatorname{max}\bigl({\tau}_r(1),\, {\tau}_r(L-1) \bigr) + \epsilon,
 \label{interval_Tr3}
\end{align}
which can be understood as an extension of \eqref{interval_Trl}.\footnote{Note that if $\Delta_r<2\epsilon$, we find that $\mathcal{T}_r = \bigcup_{l=1}^{L-1}\mathcal{T}_r(l)$.}
Based on this, we define the set similar to \eqref{subset} without the dependency on $l$:
 \begin{align}
\Aboxed{\tilde{\mathcal{R}}(n) 
&= \left\{r \in \mathcal{R}\,\left|\, n \in \frac{\mathcal{T}_r}{T_s} \right.\right\}}
 \label{subset2}
 \end{align}
The approximation in \eqref{h2h1} can then be replaced by
 \begin{equation}
 h(l,n) \approx \sum_{n'} \sum_{r\in \tilde{\mathcal{R}}(n)}  \sinc \left( n- \frac{\Delta_{r}}{T_s}-n' \right)  h(l-1, n') - e(l,n)
 \label{h2h1_inv}
\end{equation}
with the error term $e(l,n)$ redefined as
\begin{align}
e(l,n) &= \sum_{n'}
\sum_{\substack{r \in \tilde{\mathcal{R}}(n)\\ r' \in \mathcal{R} \\ r' \neq r}} 
\sinc\left( n- \frac{\Delta_{r}}{T_s}-n'\right)\sinc\left (n'-\frac{\tau_{r'}(l-1)}{T_s}\right)
\end{align}
Again, the product $\sinc( n- \Delta_{r}/{T_s}-n')\sinc (n'-({\tau_{r'}(l-1)}/{T_s})$
will be negligible if the two sinc-functions are misaligned, i.e. if $|nT_s-{\Delta_{r}} - {\tau_{r'}(l-1)}| > \epsilon$. 
As shown in the Appendix B, this condition is equivalent to
\begin{align}
\mathcal{T}_{r'}^{\scriptscriptstyle{-}} \cap \mathcal{T}_{r}^{\scriptscriptstyle{-}} &= \varnothing, \quad \quad r' \neq r \label{conditionLocInvariant}
\end{align}
where
\begin{align} 
\mathcal{T}_r^{\scriptscriptstyle{-}} &= \mathcal{T}_r-\Delta_r\nonumber\\
 &= [{\tau}_{r|\operatorname{min}}^{\scriptscriptstyle{-}},\, {\tau}_{r|\operatorname{max}}^{\scriptscriptstyle{-}}],  \label{interval_Trmin1}\\
\text{with} \quad {\tau}_{r|\operatorname{min}}^{\scriptscriptstyle{-}} &= {\tau}_{r|\operatorname{min}} - \Delta_r = \operatorname{min}\bigl({\tau}_r(0),\, {\tau}_r(L-2) \bigr) -\epsilon - \Delta_r,\label{interval_Trmin2}\\
{\tau}_{r|\operatorname{max}}^{\scriptscriptstyle{-}} &= 
{\tau}_{r|\operatorname{max}} - \Delta_r = \operatorname{max}\bigl({\tau}_r(0),\, {\tau}_r(L-2) \bigr) + \epsilon - \Delta_r, \label{interval_Trmin3}
\end{align}
and the limits in \eqref{interval_Trmin2}--\eqref{interval_Trmin3} are obtained from \eqref{interval_Tr2}--\eqref{interval_Tr3} and
${\tau}_r(1) - \Delta_r = {\tau}_r(0)$ and ${\tau}_r(L-1) - \Delta_r = {\tau}_r(L-2)$ according to \eqref{linearTOAs}.
The condition in \eqref{conditionLocInvariant} essentially states that the intervals that the individual reflections occupy over the range of the trajectory  (with exclusion of the end point $l = L-1$) may not overlap.
This can be understood as an extension of condition \eqref{assumption2} derived for the location-variant transition matrix.
Neglecting $e(l,n)$ and referring back to the vector representation of RIRs in \eqref{h_vec} it can be seen that \eqref{h2h1_inv} can be expressed in the form $\bh(l) \approx \bA \bh(l-1)$ as anticipated in \eqref{tranConst}, where the element of $\bA$ at index $(n+1, n'+1)$ is given by
\begin{equation}
\boxed{A_{n+1,n'+1}= \sum_{r\in \tilde{\mathcal{R}}(n)} \sinc \left( n- \frac{\Delta_{r}}{T_s}-n' \right).}
\label{elements_transition_matrix_invariant}
\end{equation}
if $\tilde{\mathcal{R}}(n) \neq \varnothing$ and otherwise $A_{n+1,n'+1}(l) = 0$.

Given the assumption of access to exact reflection TOAs $\tau_r(0)$ and $\tau_r(L-1)$, we estimate ${\Delta}_r$ using Equation \eqref{linearTOAs} as follows:
\begin{align}
{\Delta}_r = \frac{{\tau}_r(L-1) - {\tau}_r(0)}{L-1},
\label{estimatedTDOAs}
\end{align}
Note that from a conceptual point of view, the terms ${\tau}_r(l)$, ${\tau}_r(0)$, ${\tau}_r(L-1)-{\tau}_r(0)$, and $l/(L-1)$ in \eqref{linearTOAs} and \eqref{estimatedTDOAs} are comparable to  
$\bn_r(l)$, $\bn_r(0)$, $\bn_r(L-1) - \bn_r(0)$, and $\beta(l)$ in  \eqref{sample_based_interpolation}, respectively. 
Note, however, that the latter are sample-based, while the former are defined in continuous time-shifts and can indeed be used to implement non-integer shifts through the use of sinc-functions in \eqref{elements_transition_matrix}.

\begin{figure}[ht]
\begin{center}
\includegraphics[width=14cm]{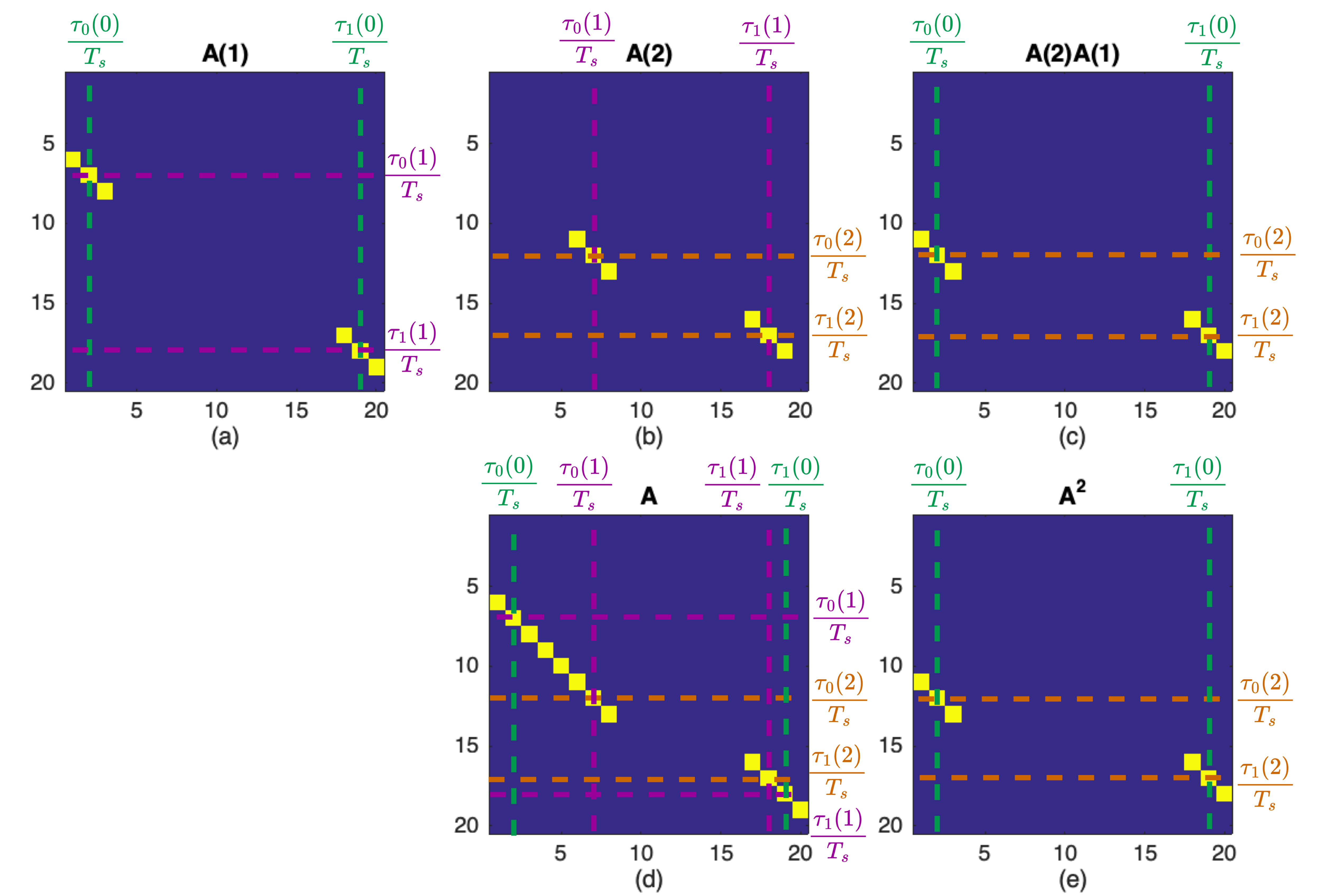}%
\end{center}
\caption{Illustration of the analytical transition matrices for a toy problem along a trajectory with $ L = 3 $, with $ \frac{2\epsilon}{T_s} = 3 $ and integer reflection TOAs. Subfigures \textbf{(a)} and \textbf{(b)} show the location-variant matrices $ \mathbf{A}(l) $ for $ l = 1 $ and $ l = 2 $, respectively. Subfigure \textbf{(c)} shows their product, $ \mathbf{A}(2) \mathbf{A}(1) $. Subfigure \textbf{(d)} shows the location-invariant matrix $ \mathbf{A} $, while \textbf{(e)} depicts $ \mathbf{A}^2 $ (i.e., $ \mathbf{A}^{L-1} $ for $ L = 3 $).}\label{fig:toyp1}
\end{figure}

Keeping in mind Figure \ref{fig:traj2} (a) and (b), for an intuitive understanding of the relationship between $\bA(l)$ and $\bA$, we refer the reader to the toy example illustrated in Figure \ref{fig:toyp1}.
Here a trajectory with $L=3$ is considered, and matrices $\bA(l)$ and $\bA$ are respectively formed using equations \eqref{elements_transition_matrix} and \eqref{elements_transition_matrix_invariant} with integer TOAs and $2\epsilon/T_s=3$. It can be seen that the result of taking $\bA^{L-1}$ yields an identical matrix to that of the multiplication of $\bA(L-1)\cdots\bA(2)\bA(1)$ (see Figure \ref{fig:toyp1} (c),(e)). Furthermore we note that the resulting matrix is equivalent to the matrix constructed using either \eqref{elements_transition_matrix} or \eqref{elements_transition_matrix_invariant} for the case of $L=2$, i.e. a transition matrix between the start and end points of the trajectory as shown in Figure \ref{fig:traj2} (c).

\subsection{Dynamic Time Warping Transition Matrices}  \label{dtw}
As it stands, the analytical solution proposed in Section \ref{locationInvAnalytical} cannot be applied directly to RIRs and requires inherent knowledge of the TOAs in $\bh(0)$ and $\bh(L-1)$. Therefore, we propose an approach to estimate the matrix $\mathbf{A}$ using a DTW algorithm to temporally align elements in $\mathbf{h}(0)$ and $\mathbf{h}(L-1)$.
DTW achieves temporal alignment by flexibly warping time, ensuring that similar patterns in the two sequences are matched, despite variations in timing. As discussed in Section \ref{section_SOTA}, prior applications of DTW have involved linear interpolation between two impulse responses. However, the method proposed here builds upon the observation that the warp path derived from DTW can be linked to both the set $\tilde{\mathcal{R}}(n)$ and $\tau_r(L-1) - \tau_r(0)$, the latter of which is necessary for estimating $\Delta_r$ via Equation \eqref{estimatedTDOAs}. Leveraging these parameters enables the construction of an appropriate analytical location-invariant transition matrix $\bA$, denoted as $\hat{\bA}_{\operatorname{dtw}}$.

\begin{figure}
\begin{center}
\includegraphics[width=15cm]{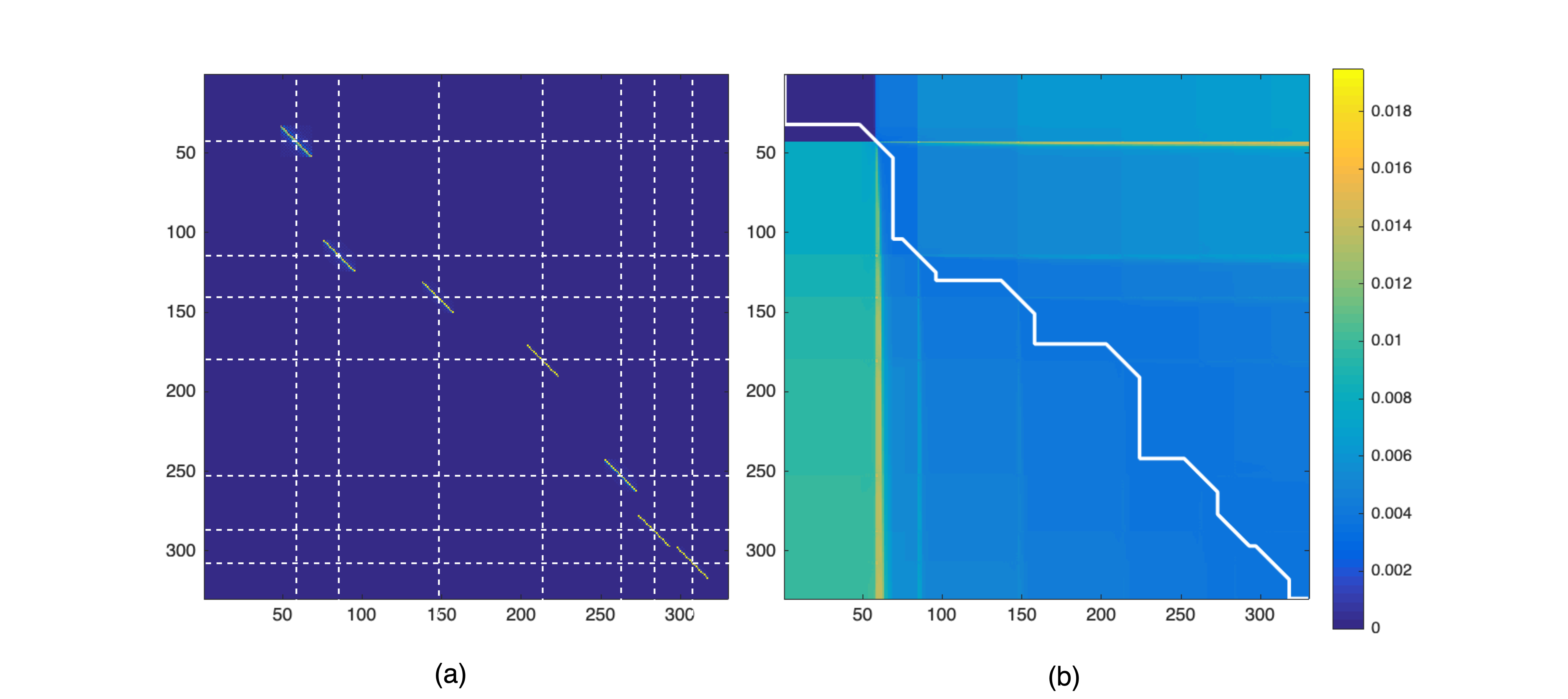}%
\end{center}
\caption{\textbf{(a)} Location-invariant matrix $\bA$ for the special case of $L=2$  and \textbf{(b)} Accumulated distance matrix $\bD$ constructed from $\bh(0)$ and $\bh(L-1)$ over-layed with DTW warp path (white line).}\label{fig:transition matrix}
\end{figure}

The DTW algorithm \citep{muller2007dynamic} involves the computation of an accumulated cost matrix $\bD \in \mathbb{R}^{(N+1) \times (N+1)}$, representing the cumulative costs of aligning each element in $\bh(L-1)$ with each element in $\bh(0)$, respectively indexed by $n$ and $n'$. Here, the cost is defined as the Euclidean distance $\lVert h(L-1,n)-h(0,n')\rVert_2$, and $\bD$ is constructed using the recurrence relation
\begin{equation}
D_{n+2, n'+2} = \lVert h(L-1,n)-h(0,n')\rVert_2 + \min \{D_{n+1, n'+2}, D_{n+2, n'+1}, D_{n+1, n'+1}\}, 
\end{equation}
which is initialized such that $D_{1,1}=0$, while $D_{n+2,1}$ and $D_{1,n'+2}$ are set to $\infty$ for all $n$ and $n'$ respectively. \footnote{In this paper, matrices are indexed from (1,1), following the convention of many programming languages. However, our RIR sequences are indexed starting from 0. To align with the zero-based indexing of our RIRs and accommodate the additional row and column used solely for initialization, we index the entries of the accumulated cost matrix as $ D_{n+2, n'+2}$}. Owing to the prominent reflection peaks in the structure of early RIRs, the matrix $\bD$ will be formed with grid-like entries (see Figure \ref{fig:transition matrix} (b))

A warp path, denoted by pairs of indices $(n+2, n'+2)$ representing optimal alignment in the sense of smallest accumulated cost, is established by backtracking from $D_{N+2,N+2}$ to $D_{2,2}$. The recurrence relation guarantees a monotonically increasing order, thus ensuring that peaks in sequences are mapped in the same order they appear, preserving temporal correspondence during alignment. Note, a single point in one sequence can correspond to multiple points in the other sequence, and vice versa. An illustration of the warp path through $\bD$ can be seen in Figure \ref{fig:transition matrix} (b).

The elements along the horizontal and vertical `grid-lines' in $\bD$ contain local maxima values, whose row and column indices should respectively correspond to integer estimates of reflection TOA time-shift indices $\frac{\tau_r(L-1)}{T_s}$ and $\frac{\tau_r(0)}{T_s}$. The elements at the intersections of these `grid-lines' represent local minima in $\bD$, signifying potential alignment points between reflection peaks in $\mathbf{h}(L-1)$ and $\mathbf{h}(0)$. Consequently, the warp path tends to traverse these intersections in a diagonal manner, guided by the minimization of alignment costs. Here a clear link between the shape of the warp path through $\bD$ (Figure \ref{fig:transition matrix} (b)) and the analytical matrix $\bA$ constructed using \eqref{elements_transition_matrix_invariant} for $L=2$ (Figure \ref{fig:transition matrix} (a)) becomes apparent. Recall that the case of $L=2$ results in a matrix $\bA$ that maps the RIR at the start of the trajectory to the RIR at the end of the trajectory (see Figure \ref{fig:traj2} (c)).

\begin{figure}[ht]
\begin{center}
\includegraphics[width=14cm]{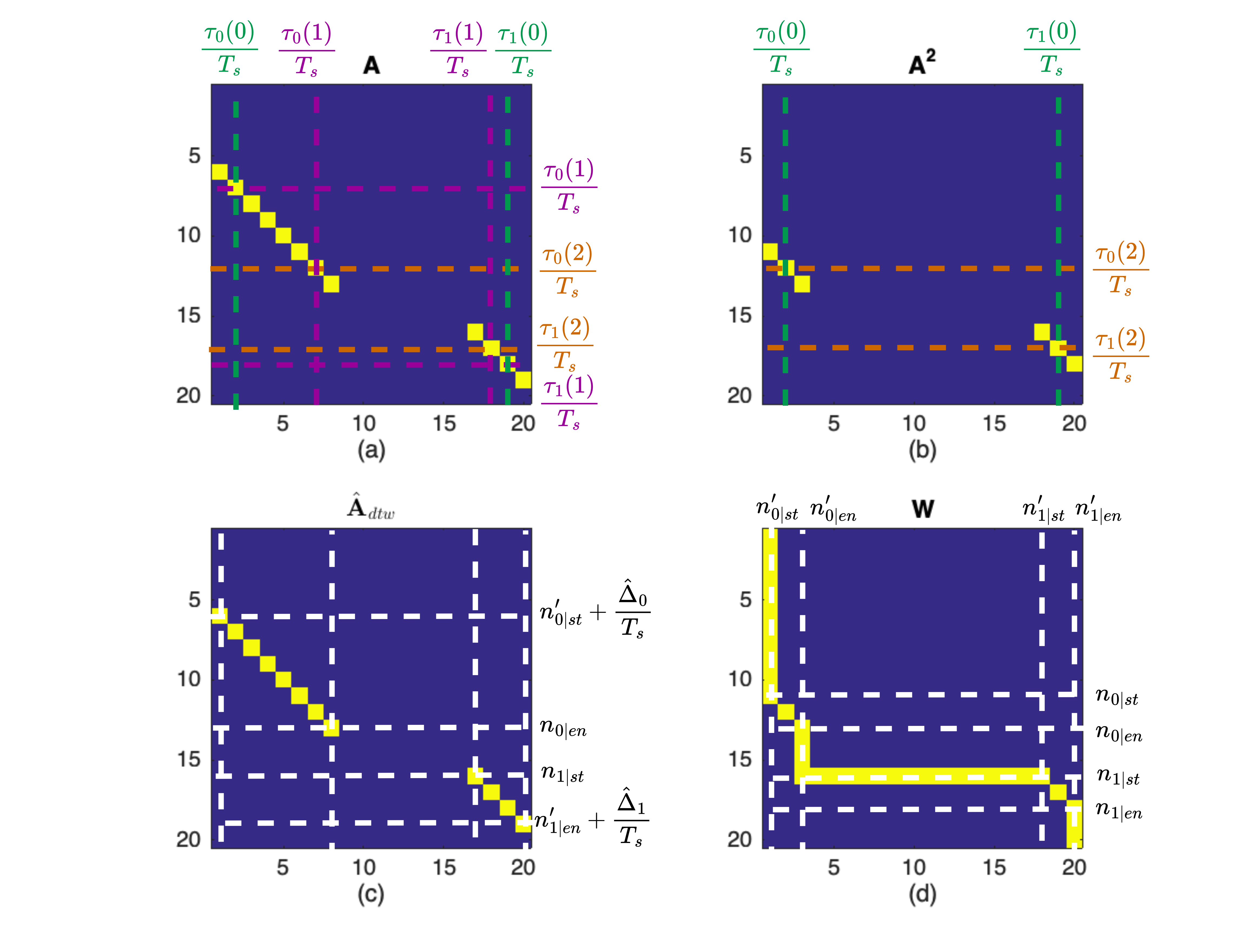}%
\end{center}
\caption{Illustration of the analytical transition matrices for a toy problem along a trajectory with $ L = 3 $, with $ \frac{2\epsilon}{T_s} = 3 $ and integer reflection TOAs. Subfigure \textbf{(a)} shows the location-invariant matrix $ \mathbf{A} $, while \textbf{(b)} depicts $ \mathbf{A}^2 $ (i.e., $ \mathbf{A}^{L-1} $ for $ L = 3 $). Subfigures \textbf{(c)} and \textbf{(d)} respectively present the matrix $ \mathbf{W} $, derived from the DTW warp path, and the estimated location-invariant matrix $ \hat{\mathbf{A}}_{\operatorname{dtw}} $.}\label{fig:toyp2}
\end{figure}

In order to use the shape of the warp path through $\bD$, we need to construct a matrix \(\bW\in \mathbb{R}^{N \times N}\), with ones placed at the indices corresponding to the pairs in the warp path and zeros elsewhere, i.e.
 \[ W_{n+1, n'+1} = \begin{cases} 1 & \text{if } (n+2, n'+2) \text{ is in the warp path} \\ 0 & \text{otherwise} \end{cases} \]
Such a matrix $\bW$ is shown in Figure \ref{fig:toyp2} (d) for the previous toy problem with integer TOAs and $\frac{2\epsilon}{T_s}=3$. It can be seen that the diagonal segments of $\bW$ are consistent with the toy problem matrix $\bA^{L-1}$, which is again displayed in Figure \ref{fig:toyp2} (b). We therefore propose to leverage the positions of the diagonal segments in $\bW$ to obtain the estimates of $\Delta_r$ and $\tilde{\mathcal{R}}(n)$ needed to construct $\bA$ as follows.

Firstly, integer estimates of $(\tau_r(L-1)-\tau_r(0))/T_s$ are obtained by measuring the `distances' (number of elements) from the diagonal segments in $\bW$ to the main diagonal. These estimates can in turn be used to find estimates of $\Delta_r$ using the relation in \eqref{estimatedTDOAs} and the known sampling period $T_s$. We employ $\hat{\Delta}_r$ and $\hat{\tau}_r$ to distinguish the estimates found in this manner, i.e.
\begin{equation}
{\hat{\Delta}}_r = \frac{{\hat{\tau}}_r(L-1) - {\hat{\tau}}_r(0)}{L-1},
\label{eqyref}
\end{equation}
Secondly, the start and end indices of the diagonal segments, denoted respectively as $(n_{r|\text{st}}, n_{r|\text{st}}')$ and $(n_{r|\text{en}}, n_{r|\text{en}}')$, are retrieved. These indices are then used to find a suitable estimate of the range $\mathcal{T}_r$ for the set $\tilde{\mathcal{R}}(n)$ in Equation \eqref{subset2} as,
\begin{align}
\hat{\mathcal{T}_r} &= [\hat{{\tau}}_{r|\operatorname{min}},\, \hat{{\tau}}_{r|\operatorname{max}}]  \label{interval_dtwTr1}\\
 \hat{{\tau}}_{r|\operatorname{min}} &= T_s\cdot \operatorname{min}\left(n'_{r|\text{st}} + \frac{\hat{\Delta}_r}{T_s},\, n_{r|\text{st}} \right),  \label{interval_dtwTr2}\\
\hat{{\tau}}_{r|\operatorname{max}} &= T_s \cdot \operatorname{max}\left(n_{r|\text{en}},\, n'_{r|\text{en}} + \frac{\hat{\Delta}_r}{T_s} \right),
 \label{interval_dtwTr3}
\end{align}
It should be ensured that the estimated ranges still result in the condition in Equation \eqref{conditionLocInvariant} being met, i.e. corresponding column ranges do not overlap.
Finally, we construct a matrix $\hat{\bA}_{\operatorname{dtw}}$ using Equations \eqref{elements_transition_matrix_invariant} and \eqref{subset2} with $\Delta_r = \hat{\Delta}_r$ and $\mathcal{T}_r=\hat{\mathcal{T}_r}$, where $\hat{\Delta r}$ and $\hat{T}_r$ are obtained using Equations \eqref{eqyref} to \eqref{interval_dtwTr3}. Figure \ref{fig:toyp2} (c) illustrates an example of a matrix obtained using this method for the previous toy problem. It is apparent that $\hat{\bA}_{\operatorname{dtw}}$ closely aligns with the desired $\bA$ matrix shown in Figure \ref{fig:toyp2} (a). Importantly, the described method allows us to derive $\bA$ without needing to explicitly estimate individual TOAs.
\section{Simulations}\label{simulations}
In this section, we systematically evaluate the performance of transition matrices $\bA$ and $\hat{\bA}_{\operatorname{dtw}}$ in State-Space time-variant RIR estimation. The algorithms considered are summarized in Table \ref{tab:algorithms}, with $\textbf{LI-}\bA$ and $\textbf{KF-}\alpha$ serving as reference methods, and $\textbf{KF-}\bA$ and $\textbf{KF-}\hat{\bA}_{\operatorname{dtw}}$ representing our primary contributions from the proposed methods outlined in Section \ref{proposedapproach}. In the following sections, we provide details on the experimental setup, including the simulated acoustic environment, Kalman filter parameters, and the performance measure used. The results of the experiments are presented in Section \ref{results}.
\begin{table}[h]
    \centering
    \begin{tabular}{|c|p{0.6\linewidth}|}
        \hline
        \rule{0pt}{13pt}\textbf{Algorithm Name} & \textbf{Algorithm Description} \\
        \hline
        \rule{0pt}{13pt} $\textbf{LI-}\bA$ & Linear interpolation equivalent to Kalman Filter update equations \eqref{eq:state-prediction}-\eqref{eq:covariance-update} with the analytical transition matrix $\bA$ derived in Section \ref{locationInvAnalytical} and $y_{\Omega}(l)=0$ and $\bx_{\Omega}(l)=0$ \\[8pt]
        \hline
        \rule{0pt}{13pt} $\textbf{KF-}\alpha$ & Kalman Filter update equations \eqref{eq:state-prediction}-\eqref{eq:covariance-update} with $\bA$ replaced with $\alpha=1$. \\[8pt]
        \hline
        \rule{0pt}{13pt} $\textbf{KF-}\bA$ & Kalman Filter update equations \eqref{eq:state-prediction}-\eqref{eq:covariance-update} with the analytical transition matrix $\bA$ derived in Section \ref{locationInvAnalytical}  \\[8pt]
        \hline
        \rule{0pt}{13pt} $\textbf{KF-}\hat{\bA}_{\operatorname{dtw}}$ & Kalman Filter update equations \eqref{eq:state-prediction}-\eqref{eq:covariance-update} with an estimate $\hat{\bA}_{\operatorname{dtw}}$ of the analytical matrix $\bA$ as derived in Section \ref{dtw} \\[8pt]
        \hline
    \end{tabular}
    \caption{Description of the Algorithms compared in the experimental results.}
    \label{tab:algorithms}
\end{table}

\subsection{Acoustic Environment}
\begin{figure}[ht]
\begin{center}
\includegraphics[width=7cm]{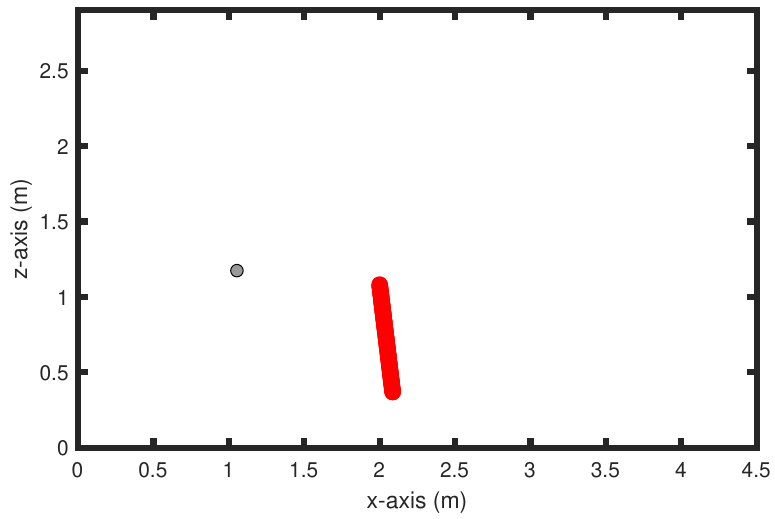}%
\end{center}
\caption{2-D illustration depicting the source position and the trajectory of the microphone within a box-shaped room, as utilized for simulations.}
\label{fig:config}
\end{figure}

For simplicity, we consider a box-shaped room with dimensions bounded by $[0, 4.50] \times [0, 5.80] \times [0, 2.90]$ m. We place a fixed-position sound source at $[1.05, 2.98, 1.17]$ m, and define a linear trajectory for the microphone spanning approximately 0.7 m, starting at $[1.94, 3.10, 1.09]$ m and ending at $[1.99, 2.95, 0.37]$ m. Along this trajectory we have established that all first-order reflections arrive in the same order. The set-up is shown in Figure \ref{fig:config}.

At a standard speech sampling rate of $16$ kHz  \citep{Nophut2024} and with a microphone velocity of $0.25$ m/s, the ISM \citep{allen1979image} is used to simulate the ground-truth early RIRs $\bh(l)$, including reflections up to the second order, at various points along the trajectory. Subsequently, the observed output signal $y_{\Omega}(l)$ is obtained using \eqref{observation_equation_prop} for a given input $x_{\Omega}(l)$ and measurement noise $v_{\Omega}(l) \sim \mathcal{N}(0, \sigma_y^2)$, where the variance $\sigma_v^2$ is adjusted based on the experimental conditions. White Gaussian noise is chosen as the input signal, denoted as $x(l) \sim \mathcal{N}(0, \sigma_x^2)$, with the variance $\sigma_x^2$ set as \(10 \log_{10}(\sigma_x^2) = -20\) dB.

\subsection{Kalman Filter Parameters}
As outlined in Section \ref{sec_update_equations}, the Kalman filter relies on specific input parameters for its operation. These parameters include the process noise covariance matrix $\mathbf{Q}$ and the variance of the measurement noise $R$. Additionally, the filter requires initialization with a state estimate $\hat{\mathbf{h}}^{+}(0)$ and an initial guess for the state estimation error covariance matrix $\mathbf{P}^{+}(0)$.

Given the assumption of knowledge of the RIR at the start of the trajectory $\mathbf{h}(0)$, the filter is initialized with $\hat{\mathbf{h}}^{+}(0) = \mathbf{h}(0)$. Since we are simulating measurement noise with variance $\sigma_v^2$, we set $R = \sigma_v^2$. As previously mentioned, in practice $R$ can potentially be estimated using background noise recordings. The appropriate $\bA$ matrix is selected for each algorithm, as outlined in Table \ref{tab:algorithms} with the considered width $2\epsilon$ of the $\sinc$ functions in the proposed analytical transition matrices set to $\frac{2\epsilon}{T_s}=20$ . It was found that for the reference Kalman filter, a transition factor of $\alpha=1$ was suitable and varying this did not change the results in a significant manner. After some initial testing of the Kalman filter, the tuning parameter $\mathbf{Q}$ was chosen as $\mathbf{Q} = \sigma_w^2 \mathbf{I}$, where $\bI$ is an identity matrix and  $10\log(\sigma_w^2) = -30$ dB.

\subsection{Performance Measure}
In assessing the accuracy of our estimated RIRs $\hat{\mathbf{h}}^{+}(l)$, we employ the normalized misalignment $\mathcal{M}_{\operatorname{dB}}(l)$. This metric is defined as follows:
\begin{equation}
\mathcal{M}_{\operatorname{dB}}(l)= 20 \log_{10}\left(\frac{\lVert \hat{\mathbf{h}}^{+}(l) - \mathbf{h}(l) \rVert_2}{\lVert \mathbf{h}(l) \rVert_2}\right)
\end{equation}
Here, $\mathbf{h}(l)$ represents the ground truth early RIRs. 

\subsection{Experiments}\label{exp}
We categorize experiments into the following four sets to evaluate algorithm performance.

\subsubsection{Experiment 1 -- Ideal case} In the first set of experiments, we examine the `ideal' scenario where no noise is present on the observed output signal $y_{\Omega}(l)$  ($\sigma_v^2=0$) and every sample along the trajectory is considered ($\Omega=1$). Simulated RIRs exclusively include first-order reflections.
\subsubsection{Experiment 2 -- Noise sensitivity} Building upon the setup in \textbf{Experiment 1} with $\Omega=1$ and only first-order reflections included, the second set of experiments evaluates algorithm robustness when different noise levels are added to the observed output signal, i.e. varying $\text{SNR}=\{6,0,-6\}$ dB.
\subsubsection{Experiment 3 -- Spatial sampling effects} The third set of experiments employs different spatial downsampling factors $\Omega$ to assess the impact of using fewer data points along the trajectory. In this case we once again consider $\sigma_v^2=0$ and only first-order reflections, while varying $\Omega =\{2,8,32\}$. Note that $\Omega=2$ corresponds to $50\%$ of the samples along the trajectory being used.
\subsubsection{Experiment 4 -- Second-order reflections included} The final set of experiments investigates the inclusion of second-order reflections in the RIRs and the corresponding simulated observed signal. We once again examine the `ideal' scenario where no noise is present on the observed output signal $y_{\Omega}(l)$ ($\sigma_v^2=0$) and every sample along the trajectory is considered ($\Omega=1$). Please note that the analytical transition matrix $\bA$ is still constructed solely using first-order TOAs, as not all second-order reflections meet the necessary assumptions along the given trajectory. As a robustness measure, after constructing $\bA$, ones are added along the main diagonal in any part of the matrix with empty rows. This step ensures that second-order reflections are not entirely discarded. On the other hand, the DTW algorithm operates directly on the RIRs and thus, when constructing $\hat{\mathbf{A}}_{\text{dtw}}$, it inherently assumes that all reflections arrive in the same order in $\mathbf{h}(0)$ and $\mathbf{h}(L-1)$. In other words, it may incorrectly associate some reflections in $\mathbf{h}(0)$ and $\mathbf{h}(L-1)$ as belonging to the same image source.

\section{Results}\label{results}
We present the results corresponding to the experiments outlined in Section \ref{exp} above. The reader is urged to take careful note of the different $y$-axis scaling in the presented results.

\subsubsection{Result 1: Ideal Case}
\begin{figure}[ht]
\begin{center}
\includegraphics[width=13cm]{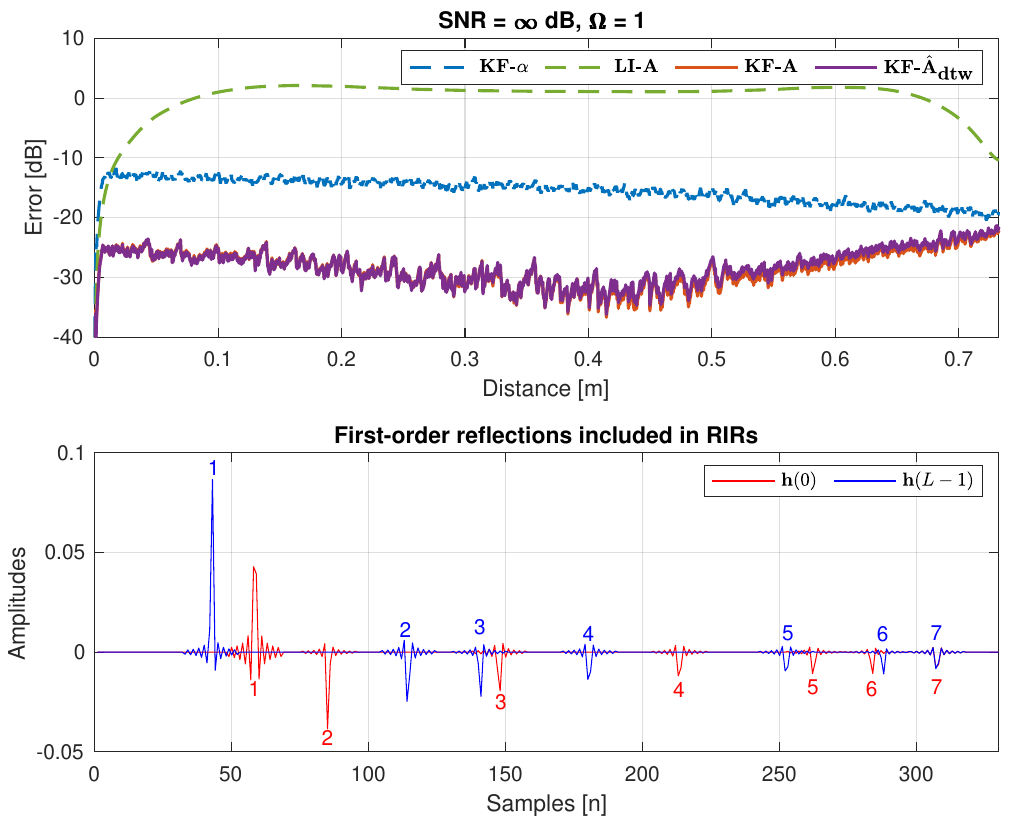}%
\end{center}
\caption{(Top) Result 1-- Normalized misalignment along the trajectory between the `ground-truth' RIR $\bh(l)$ and the estimated RIR $\hat{\bh}(l)$ using algorithms detailed in Table \ref{tab:algorithms}. This is for the case of a noiseless observed output signal $y_{\Omega}(l)$. (Bottom) Simulated RIRs corresponding to the start and end of the trajectory, denoted as $\bh(0)$ and $\bh(L-1)$, respectively. Peaks corresponding to the same source or reflection are labeled by the same number in each RIR.}\label{fig:result1}
\end{figure}

The results of \textbf{Experiment 1} are presented at the top of Figure \ref{fig:result1}, where the normalized misalignment $\mathcal{M}_{\operatorname{dB}}(l)$ between the ground-truth RIR $\bh(l)$ and the estimated RIR $\hat{\bh}(l)$ is displayed for the various algorithms at different positions along the trajectory. The bottom of Figure \ref{fig:result1} shows the start and end point RIRs used to obtain our transition matrices. This is provided to give the reader an idea of how the source and reflection TOAs change over the trajectory. Furthermore, from the labeled TOAs it can be seen that there are no overlapping intervals, which aligns with our assumptions.

At the beginning of the trajectory, all algorithms exhibit zero error, owing to their initialization with a known RIR $\bh(0)$. Following a subsequent steep increase in error, distinctive properties of the curves resulting from the different algorithms become evident.
The reference Kalman filter algorithm ($\textbf{KF-}\alpha$), which utilizes a transition factor of $\alpha=1$ and relies solely on microphone observations, demonstrates a gradual decrease in error throughout the remainder of the trajectory. Excluding the start and end segments, it can be observed that Algorithm $\textbf{LI-}\bA$, whose performance solely relies on the model of $\bA$, results in an error curve with a subtle dip in it. This non-monotonic behavior can be attributed to the position of the direct source relative to the linear microphone trajectory (see Figure \ref{fig:config}) and the fact that the source is not sufficiently in the far-field for our assumption that TDOAs remain approximately constant to hold. Towards the end of the trajectory, Algorithm $\textbf{LI-}\bA$ exhibits a rapid decrease in error. This is because the matrix $\bA$ is derived from accurate TOAs at the start and end of the trajectory, and therefore estimates should most closely align with the ground-truth RIRs around these positions. 

In general, changes in the error curve of the Kalman filter algorithms using a transition matrix (\(\textbf{KF-}\bA\) and \(\textbf{KF-}\hat{\bA}_{\operatorname{dtw}}\)) correlate with changes in the error curve of Algorithm \(\textbf{LI-}\bA\). However, a more significant dip in error is observed, and the curves nearly converge with those of Algorithm \(\textbf{KF-}\alpha\) at the end of the trajectory. This behavior may be a result of the rapid convergence of \(\textbf{KF-}\bA\) and \(\textbf{KF-}\hat{\bA}_{\operatorname{dtw}}\) under ideal conditions (i.e., no noise), making the Kalman filter more sensitive to the degradation of the transition matrix model accuracy.

If we now consider the relative performance of the algorithms, we note that the Kalman filter employing the `ideal' analytical transition matrix, Algorithm $\textbf{KF-}\bA$, showcases a substantial improvement of on average approximately $-15$ dB compared to the reference Kalman filter, Algorithm $\textbf{KF-}\alpha$. Importantly, similar improvement extends to the case where the analytical transition matrix approximated using DTW is used i.e, Algorithm $\textbf{KF-}\hat{\bA}_{\operatorname{dtw}}$. Moreover, the advantage of employing a state-space model over a linear interpolation method becomes evident from the comparatively poor overall results of Algorithm $\textbf{LI-}\bA$. The observed enhancements in accuracy lend support to our preliminary hypothesis: the combination of room acoustic model-based interpolation and a data-driven state-space model provides a more accurate approach to early RIR estimation.

\subsubsection{Result 2: Noise Sensitivity}
\begin{figure}[ht]
\begin{center}
\includegraphics[width=12.5cm]{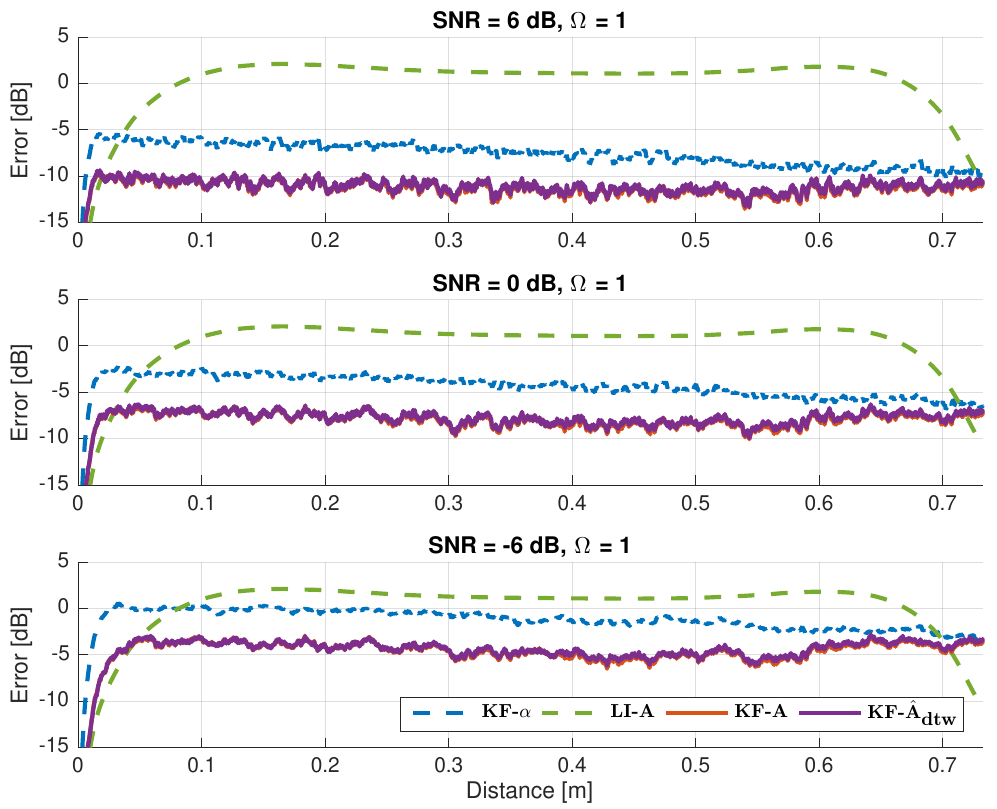}%
\end{center}
\caption{Result 2-- Normalized misalignment along the trajectory between the `ground-truth' RIR $\bh(l)$ and the estimated RIR $\hat{\bh}(l)$ using algorithms detailed in Table \ref{tab:algorithms}. This assessment is conducted for different levels of noise added to the observed output signal $y_{\Omega}(l)$.}\label{fig:result2}
\end{figure}

The results of \textbf{Experiment 2} are presented in Figure \ref{fig:result2} and once again exhibit error curve shapes consistent with those observed in \textbf{Experiment 1}. However, as anticipated, overall performance is worse, except for Algorithm $\textbf{LI-}\bA$, which remains unaffected since it does not rely on the noisy microphone measurements. Additionally, we observe slower convergence and less pronounced dips in the error curves of the algorithms \(\textbf{KF-}\bA\) and \(\textbf{KF-}\hat{\bA}_{\operatorname{dtw}}\) compared to the noiseless case. This is likely because the less reliable observations prevent convergence to such low error levels in the first place, making the dips less noticeable.

It is evident that even under poor SNR conditions, our methods consistently outperform both of the reference algorithms. Notably, at an SNR of $-6$ dB, an average of approximately $-5$ dB difference persists, further underscoring the advantages of the proposed approach, even when measurements contain high levels of noise.

\subsubsection{Result 3: Spatial Sampling Effects}
\begin{figure}[ht]
\begin{center}
\includegraphics[width=14cm]{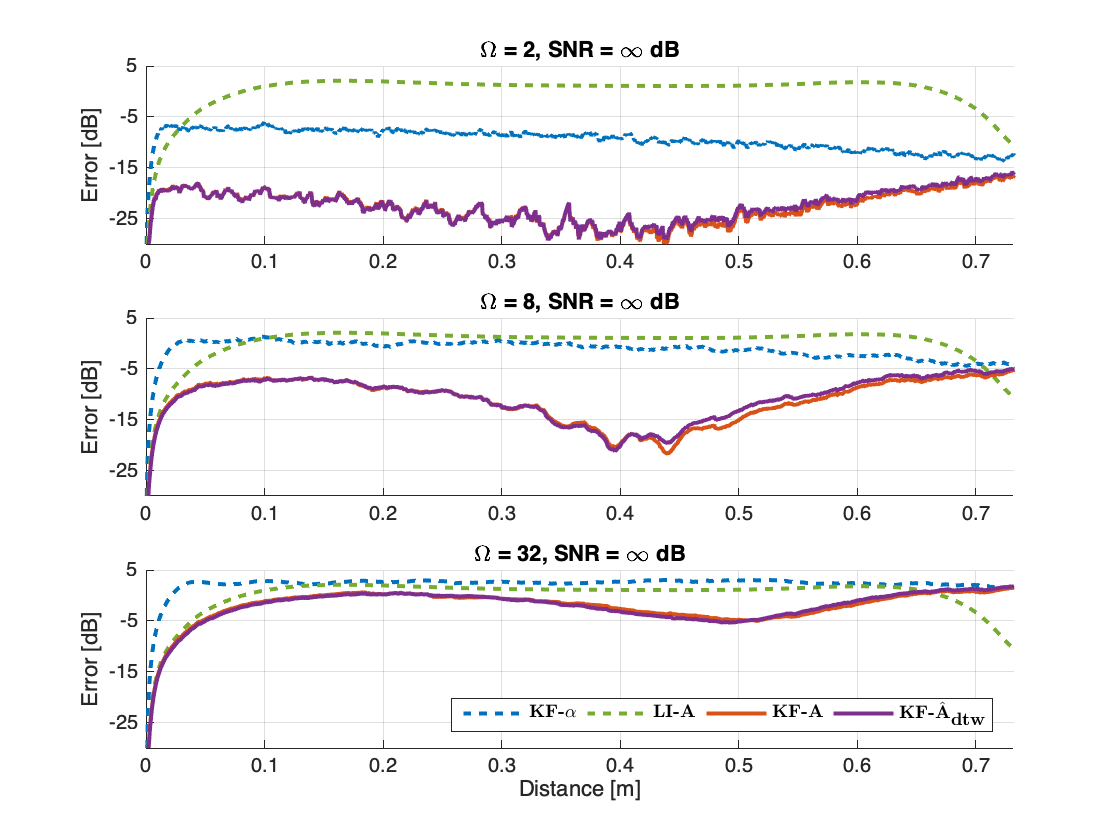}%
\end{center}
\caption{Result 3-- Normalized misalignment along the trajectory between the `ground-truth' RIR $\bh(l)$ and the estimated RIR $\hat{\bh}(l)$ using algorithms detailed in Table \ref{tab:algorithms}. This is for the case of a noiseless observed output signal $y_{\Omega}(l)$ with different spatial sampling along the trajectory as a function of $\Omega$.}
\label{fig:result3}
\end{figure}

The results of \textbf{Experiment 3} are presented in Figure \ref{fig:result3}. As fewer measurements are used, the disparity between the results obtained using the linear interpolation method and those obtained using a Kalman filter naturally diminishes. Additionally, the performance of the reference Kalman filter ($\textbf{KF-}\alpha$) deteriorates more rapidly than that of our proposed methods ($\textbf{KF-}\bA$ and $\textbf{KF-}\hat{\bA}_{\operatorname{dtw}}$). We further note an apparent shift in the position of the dip in the error curve of Algorithms $\textbf{KF-}\bA$ and $\textbf{KF-}\hat{\bA}_{\operatorname{dtw}}$ as fewer samples are used. This is likely a consequence of the change in how often the state equation is propagated.

For $\Omega= 8$, equivalent to approximately $12.5\%$ of the available sampling points, Algorithms $\textbf{KF-}\alpha$ and $\textbf{LI-}\bA$ exhibit similar performance, while our proposed algorithms consistently outperform both by an average of approximately $-12$ dB. Despite the trivial choice between the two reference methods when only a small number of measurements are available, as demonstrated by the case of $\Omega=32$ (equivalent to $3.13\%$ of sampling points),  our proposed method  still demonstrates an overall improvement in performance.

\subsubsection{Result 4: Second-Order Reflections Included}
\begin{figure}[ht]
\begin{center}
\includegraphics[width=13cm]{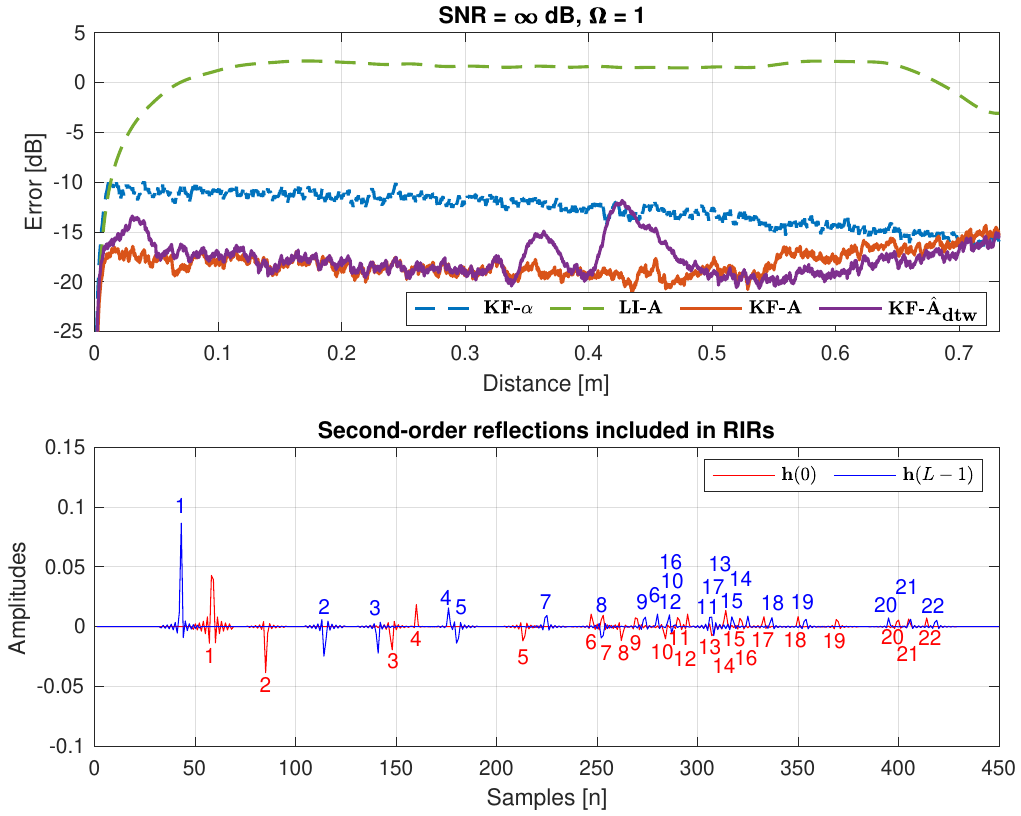}%
\end{center}
\caption{(Top) Result 4-- Normalized misalignment along the trajectory between the `ground-truth' RIR $\bh(l)$ and the estimated RIR $\hat{\bh}(l)$ using algorithms detailed in Table \ref{tab:algorithms}. This is for the case of a noiseless observed output signal $y_{\Omega}(l)$ with simulated RIRs inclusive of second-order reflections. (Bottom) Simulated RIRs corresponding to the start and end of the trajectory, denoted as $\bh(0)$ and $\bh(L-1)$, respectively. Peaks corresponding to the same source or reflection are labeled by the same number in each RIR.}\label{fig:result4}
\end{figure}

The results of \textbf{Experiment 4} are shown at the top of Figure \ref{fig:result4}. It is important to recall that in \textbf{Experiment 4}, the analytical transition matrix $\bA$ used in Algorithm $\textbf{KF-}\bA$ is still constructed using only first-order TOAs. This is because not all second-order reflections meet the necessary assumptions along the given trajectory, as indicated by the reflection TOA labels towards the end of the RIR at the bottom of Figure \ref{fig:result4}, e.g., the interval over which reflection 6 moves overlaps with reflections 7, 8, and 9. However, by adding ones along the diagonal of $\bA$ in any empty rows, we ensure that second-order reflections are not entirely discarded. This allows us to leverage prior knowledge, even if incomplete, and explains the strong convergence with Algorithm $\textbf{KF-}\alpha$ towards the end of the trajectory. That being said, Algorithm $\textbf{KF-}\bA$ still demonstrates an average improvement of approximately $-7$ dB over Algorithm $\textbf{KF-}\alpha$.

It is important to reiterate that in estimating the matrix $\hat{\bA}_{\operatorname{dtw}}$, we directly utilize the RIRs $\bh(0)$ and $\bh(L-1)$, thus implicitly incorporating second-order reflection information. However, in this case not all second-order reflections adhere to the assumption that the time intervals individual reflections occupy over the trajectory do not overlap. Given that DTW inherently maps sequences in a monotonic manner, assuming consistent order among significant points like peaks, certain reflection peaks might be incorrectly mapped between RIRs. Consequently, the points along the trajectory where second-order reflections in the ground-truth RIRs begin to overlap likely correspond to the jumps observed in the error curve of Algorithm $\textbf{KF-}\hat{\bA}_{\operatorname{dtw}}$. Nonetheless, this serves as a robust test demonstrating that despite deviations from certain assumptions regarding all reflections, the proposed estimated matrix still yields an overall enhancement over Algorithm $\textbf{KF-}\alpha$.

\newpage
\section{Conclusions}
In conclusion, this paper delves into the estimation of early segments of time-variant RIRs through a state-space model incorporating the ISM within the state transition matrix. Simulation results indicate that this approach outperforms both RIR interpolation and a purely data-driven state-space model using a transition factor. Moreover, a practical method to estimate such a matrix has been proposed and has a similar performance to the `ideal' analytical transition matrix derived. It is important to acknowledge that certain assumptions inherent to our method limit its application within specific areas of a room. This necessitates further research in order to improve the robustness of the approach, potentially through adaptive estimation of the transition matrix. Furthermore, experimental validation using real measurement data is required to assess the effectiveness of the proposed approach in real-world scenarios.

\newpage
\section*{Appendix}

\subsection*{Appendix A}

\noindent\textit{Lemma}. With definitions as in Section \ref{analy} and for $r \in \tilde{\mathcal{R}}(l,n)$, $r' \in \mathcal{R}$, and $r \neq r'$ it holds that 
\begin{equation}
\left|nT_s-{\Delta_{r}(l)} - {\tau_{r'}(l-1)}\right| > \epsilon \quad \text{if} \quad \mathcal{T}_{r'}(l-1) \cap \mathcal{T}_{r}(l-1) = \varnothing.
\label{app_start}
\end{equation}
\noindent\textit{Proof}. The inequality in \eqref{app_start} can alternatively be written as
\begin{equation}
\epsilon < nT_s-{\Delta_{r}(l)} - {\tau_{r'}(l-1)} \quad \vee \quad  nT_s-{\Delta_{r}(l)} - {\tau_{r'}(l-1)} < -\epsilon.
\label{app_start_alt}
\end{equation}
We begin by finding lower and upper bounds for $nT_s-{\Delta_{r}(l)} - {\tau_{r'}(l-1)}$ based on the constraint that $r \in \tilde{\mathcal{R}}(l,n)$. 
From the definition of $\tilde{\mathcal{R}}(l,n)$ in \eqref{subset} and with $\tau_{r}(l)-\Delta_{r}(l) = \tau_{r}(l-1)$ according to \eqref{TDOA}, we find that 
\begin{align}
&n \in \frac{\mathcal{T}_r(l)}{T_s}\nonumber\\
\Leftrightarrow {\tau_{r}(l)}  -\epsilon < \,&nT_s < {\tau_{r}(l)} + \epsilon\nonumber\\ 
\Leftrightarrow {\tau_{r}(l)}  -\epsilon - {\Delta_{r}(l)}  - {\tau_{r'}(l-1)} < \,&nT_s - {\Delta_{r}(l)}  - {\tau_{r'}(l-1)} < {\tau_{r}(l)} + \epsilon - {\Delta_{r}(l)}  - {\tau_{r'}(l-1)}\nonumber\\
\Leftrightarrow {\tau_{r}(l-1)}  -\epsilon  - {\tau_{r'}(l-1)} < \,&nT_s - {\Delta_{r}(l)}  - {\tau_{r'}(l-1)} < {\tau_{r}(l-1)} + \epsilon  - {\tau_{r'}(l-1)}.
 \label{app_set_reform}
\end{align}
We observe that \eqref{app_start_alt} is satisfied if the lower or upper bound for $nT_s-{\Delta_{r}(l)} - {\tau_{r'}(l-1)}$ in \eqref{app_set_reform} is greater than $\epsilon$ or smaller than $-\epsilon$, respectively, i.e. if 
\begin{align}
\epsilon<{\tau_{r}(l-1)} -\epsilon - {\tau_{r'}(l-1)}  \quad &\vee \quad {\tau_{r}(l-1)} + \epsilon - {\tau_{r'}(l-1)} < -\epsilon 
\nonumber\\
\Leftrightarrow \quad {\tau_{r'}(l-1)} + \epsilon<{\tau_{r}(l-1)} - \epsilon \quad &\vee \quad {\tau_{r}(l-1)} + \epsilon  < {\tau_{r'}(l-1)} - \epsilon.
\label{appA_lower_upper}
\end{align}
Since ${\tau_{r}(l-1)} - \epsilon$ and ${\tau_{r}(l-1)} + \epsilon$
are the limits of of the interval $\mathcal{T}_r(l-1)$ as defined in \eqref{interval_Trl}, the expression in \eqref{appA_lower_upper} is equivalent to 
\begin{align}
&{\tau_{r'}(l-1)} \pm \epsilon \notin \mathcal{T}_r(l-1). 
\end{align}
As ${\tau_{r'}(l-1)} \pm \epsilon$ occupies the interval $\mathcal{T}_{r'}(l-1)$, it follows that
\begin{align}
\mathcal{T}_{r'}(l-1) \cap \mathcal{T}_{r}(l-1) = \varnothing. \quad\quad\quad \square
\end{align}

\subsection*{Appendix B}

\noindent\textit{Lemma}. With definitions as in Section \ref{locationInvAnalytical} and for $r \in \tilde{\mathcal{R}}(n)$, $r' \in \mathcal{R}$, and $r \neq r'$ it holds that 
\begin{equation}
\left|nT_s-{\Delta_{r}} - {\tau_{r'}(l-1)}\right| > \epsilon \quad \text{if} \quad \mathcal{T}_{r'}^{\scriptscriptstyle -} \cap \mathcal{T}_{r}^{\scriptscriptstyle -} = \varnothing.
\label{appB_start}
\end{equation}
\textit{Proof}. The inequality in \eqref{appB_start} can alternatively be written as
\begin{align}
\Leftrightarrow \quad &\epsilon < nT_s-{\Delta_{r}} - {\tau_{r'}(l-1)} \quad \vee \quad  nT_s-{\Delta_{r}} - {\tau_{r'}(l-1)} < -\epsilon.
\label{appB_start_alt}
\end{align}
We begin by finding lower and upper bounds for $nT_s-{\Delta_{r}} - {\tau_{r'}(l-1)}$ based on the constraint that $r \in \tilde{\mathcal{R}}(n)$. 
From the definition of $\tilde{\mathcal{R}}(n)$ in \eqref{subset2}, we find that 
\begin{align}
&n \in \frac{\mathcal{T}_r}{T_s}\nonumber\\
\Leftrightarrow \quad {\tau_{r|\operatorname{min}}}  \leq\,&nT_s  \leq {\tau_{r|\operatorname{max}}} \nonumber\\
 \Leftrightarrow \quad {\tau_{r|\operatorname{min}}}-\epsilon -{\Delta_{r}} - {\tau_{r'}(l-1)} \leq \,&nT_s -{\Delta_{r}} - {\tau_{r'}(l-1)} \leq {\tau_{r|\operatorname{max}}} + \epsilon -{\Delta_{r}} - {\tau_{r'}(l-1)}.
 \label{appB_set_reform}
\end{align}
We observe that \eqref{appB_start_alt} is satisfied if the lower or the upper bound for $nT_s -{\Delta_{r}} - {\tau_{r'}(l-1)}$ on the left- and the right-hand side of \eqref{appB_set_reform} is greater than $\epsilon$ or smaller than $-\epsilon$, respectively, i.e. if 
\begin{align}
\epsilon <{\tau_{r|\operatorname{min}}} -{\Delta_{r}} - {\tau_{r'}(l-1)} \quad &\vee \quad {\tau_{r|\operatorname{max}}} -{\Delta_{r}} - {\tau_{r'}(l-1)} < -\epsilon 
\nonumber\\
\Leftrightarrow \quad {\tau_{r'}(l-1)} + \epsilon <{\tau_{r|\operatorname{min}}}-{\Delta_{r}}  \quad &\vee \quad {\tau_{r|\operatorname{max}}} -{\Delta_{r}} < {\tau_{r'}(l-1)} -\epsilon.
\label{appB_lower_upper}
\end{align}
Since ${\tau_{r|\operatorname{min}}}-{\Delta_{r}}$ and ${\tau_{r|\operatorname{max}}}-{\Delta_{r}}$
are the limits of of the interval $\mathcal{T}_r^{\scriptscriptstyle -}$ as defined in \eqref{interval_Trmin1}--\eqref{interval_Trmin3}, the expression in \eqref{appB_lower_upper} is equivalent to 
\begin{align}
&{\tau_{r'}(l-1)} \pm \epsilon \notin \mathcal{T}_r^{\scriptscriptstyle -}. 
\end{align}
As ${\tau_{r'}(l-1)} \pm \epsilon$ for $l=1,\,\dots,\,L-1$  occupies the interval $\mathcal{T}_{r'}^{\scriptscriptstyle -}$, it follows that
\begin{align}
\mathcal{T}_{r'}^{\scriptscriptstyle -} \cap \mathcal{T}_{r}^{\scriptscriptstyle -} = \varnothing. \quad\quad\quad \square
\end{align}

\section*{Funding}
This work was supported by:
\begin{itemize}
    \item The European Research Council under the European Union's Horizon 2020 research and innovation program / ERC Consolidator Grant: SONORA (no. 773268). This paper reflects only the authors' views and the Union is not liable for any use that may be made of the contained information.
    \item KU Leuven Internal Funds C14/21/075 ``A holistic approach to the design of integrated and distributed digital signal processing algorithms for audio and speech communication devices''
    \item The FWO Research Project: ``The Boundary Element Method as a State-Space Realization Problem'' (G0A0424N)
\end{itemize}

\section*{Acknowledgments}
This work has already been published in a peer-reviewed article in the journal Frontiers in Signal Processing \citep{10.3389/frsip.2024.1426082}, under the research topic ``Audio and Acoustics of Movement". The full publication can be accessed at:
\url{https://www.frontiersin.org/journals/signal-processing/articles/10.3389/frsip.2024.1426082/full}.

\newpage
\bibliography{FINAL.bib}

\end{document}